\documentclass[prd,aps,showpacs,onecolumn,longbibliography,10pt,notitlepage
              ]{revtex4-1}

\usepackage[utf8]{inputenc}

\usepackage[american]{babel}

\usepackage{graphicx}

\usepackage{amsmath}
\usepackage{amssymb}

\usepackage[colorlinks=true,
            linkcolor=blue,
            urlcolor=blue,
            citecolor=blue,
            breaklinks=true]{hyperref}

\usepackage{lmodern}
\usepackage[T1]{fontenc}

\usepackage{tensor}

\usepackage{xcolor}
\definecolor{mcolor}{rgb}{0.70,0.11,0.11}

\newcommand{\dd}{\mathrm{d}}
\newcommand{\ee}{\mathrm{e}}
\newcommand{\Tr}{\mathrm{Tr}}
\newcommand{\mcA}{\mathcal{A}}
\newcommand{\mcM}{\mathcal{M}}
\newcommand{\mcP}{\mathcal{P}}
\newcommand{\mcS}{\mathcal{S}}

\newcommand{\mcZ}{\mathcal{Z}}
\renewcommand{\Re}{\mathrm{Re}}
\renewcommand{\Im}{\mathrm{Im}}
\renewcommand*{\vec}[1]{\boldsymbol{#1}}
\renewcommand{\eqref}[1]{Eq.~(\ref{#1})}
\newcommand{\figref}[2][{}]{Fig.~\hyperref[#2]{\ref{#2}#1}}
\newcommand{\intlim}{\int\displaylimits}

\begin{document}
\title{Electron-positron annihilation into two photons in an intense plane-wave
field}

\author{S. Bragin}
\email{srg.brag@gmail.com}

\author{A. Di Piazza}
\email{dipiazza@mpi-hd.mpg.de}

\affiliation{Max Planck Institute for Nuclear Physics, Saupfercheckweg 1, 
             69117 Heidelberg, Germany}

\date{\today}

\begin{abstract}
The process of electron-positron annihilation into two photons in the presence
of an intense classical plane wave of an arbitrary shape is investigated
analytically by employing light-cone quantization and by taking into account the
effects of the plane wave exactly. We introduce a general description of
second-order \mbox{2-to-2} scattering processes in a plane-wave background
field, indicating the necessity of considering the localization of the colliding
particles and achieving that by means of wave packets. We define a local cross
section in the background field, which generalizes the vacuum cross section and
which, though not being directly an observable, allows for a comparison between
the results in the plane wave and in vacuum without relying on the shape of the
incoming wave packets. Two possible cascade or two-step channels have been
identified in the annihilation process and an alternative way of representing
the two-step and one-step contributions via a ``virtuality'' integral has been
found. Finally, we compute the total local cross section to leading order in the
coupling between the electron-positron field and the quantized photon field,
excluding the interference between the two leading-order diagrams arising from
the exchange of the two final photons, and express it in a relatively compact
form. In contrast to processes in a background field initiated by a single
particle, the pair annihilation into two photons, in fact, also occurs in
vacuum. Our result naturally embeds the vacuum part and reduces to the vacuum
expression, known in the literature, in the case of a vanishing laser field.

  
\end{abstract}
 
\maketitle

\section{Introduction}

With the development of high-power laser technology the verification of the
nonlinear-QED predictions for various phenomena in an intense background field
of a macroscopic extent is becoming attainable in laboratory experiments
\cite{marklund_nonlinear_2006, ehlotzky_fundamental_2009, di_piazza_review_2012,
narozhny_extreme_2015, king_measuring_2016}. Among QED processes in an intense
laser field, two first-order ones, nonlinear Compton scattering ($e^-
\Rightarrow e^- \gamma$) \cite{brown_interaction_1964, neville_quantum_1971,
ritus_quantum_effects_1985, ivanov_complete_2004, harvey_signatures_2009,
boca_nonlinear_2009, mackenroth_determining_2010, seipt_nonlinear_2011,
mackenroth_nonlinear_2011, krajewska_compton_2012, wistisen_interference_2014,
angioi_compton_2016, wistisen_numerical_2019} and nonlinear Breit-Wheeler pair
production ($\gamma \Rightarrow e^- e^+$) \cite{reiss_absorption_1962,
ritus_quantum_effects_1985, bulanov_multiple_2010, heinzl_finite_2010,
ipp_streaking_2011, krajewska_breit-wheeler_2012, jansen_strongly_2013,
fedotov_pair_2013, meuren_pair_creation_2015} have been extensively investigated
theoretically (see also the reviews \cite{mitter_quantum_1975,
ehlotzky_fundamental_2009, di_piazza_review_2012, narozhny_extreme_2015}), where
by a double-line arrow we highlight the fact that a process happens in a
background field, which in general has to be taken into account
nonperturbatively. Recently, nonlinear Compton scattering was also probed
experimentally and signatures of quantum effects were observed
\cite{cole_experimental_2018, poder_experimental_2018} (see
Ref.~\cite{wistisen_experimental_2018} for a related experiment carried out in
crystals). Moreover, these reactions are the only QED effects included in common
implementations of the QED particle-in-cell (PIC) scheme
\cite{arber_contemporary_2015, gonoskov_extended_2015, lobet_modeling_2016},
which is a standard tool for the numerical investigation of the interaction
between a laser field of extreme intensity ($\gtrsim 10^{23}$~W/cm$^2$) and
matter, in particular, of the dynamics of the electron-positron plasma, produced
in this interaction \cite{bell_possibility_2008, nerush_laser_2011,
ridgers_dense_2012, jirka_electron_2016, grismayer_laser_2016,
tamburini_laser-pulse-shape_2017, vranic_electron-positron_2017, luo_qed_2018,
efimenko_extreme_2018, efimenko_laser-driven_2019} (an electron-positron plasma
interacting with a background field can also arise in a collision of a
high-density electron beam with a target \cite{sarri_generation_2015} and in
some astrophysical scenarios \cite{sturrock_model_1971, usov_millisecond_1992,
wardle_electron-positron_1998, chang_longterm_2008, sinha_polarized_2019}).

Other channels of the first-order processes, i.e., electron-positron
annihilation into one photon ($e^-e^+ \Rightarrow \gamma$) and photon absorption
($e^- \gamma \Rightarrow e^-$) are typically sizable only in a small part of the
phase space of the incoming particles \cite{baier_processes_1968,
ritus_quantum_effects_1985, ilderton_pair_2011, tang_one-photon_2019}.
Therefore, if electron-positron annihilation and photon absorption are to be
also included into the consideration of the evolution of a many-particle system
in an intense laser field, which may involve different geometries of particle
collisions, it is necessary to assess the next-order processes, i.e., $e^-e^+
\Rightarrow \gamma \gamma$ and $e^- \gamma \Rightarrow e^- \gamma$,
respectively.

However, a complete evaluation of a tree-level second-order process in an
external laser field is not straightforward. For instance, first theoretical
calculations for trident process, i.e., seeded electron-positron pair production
($e^- \Rightarrow e^-e^-e^+$), were performed long ago
\cite{baier_higher-order_1972, ritus_vacuum_1972}. It was demonstrated that the
total probability can be decomposed into a two-step contribution, which is
related to the physical situation of the intermediate electron being real and
which can be reconstructed as a combination of the corresponding nonlinear
Compton and Breit-Wheeler probabilities, and a one-step contribution, for which
the intermediate electron is virtual and which was computed in part. Later,
first experiments on trident were also carried out \cite{burke_positron_1997,
bamber_studies_1999}. But only recently, via a series of works, a full
evaluation of the trident process was presented for the constant-crossed and
general plane-wave background field cases \cite{hu_trident_2010,
ilderton_trident_2011, king_trident_2013, dinu_trident_2018, king_trident_2018,
mackenroth_trident_2018} (for an estimation, the one-step part of trident is
sometimes taken into account with the use of the Weizs\"{a}cker-Williams
approximation \cite{blackburn_quantum_2014, hojbota_effect_2018}; see also Ref.
\cite{king_trident_2013}). A result for double Compton scattering ($e^-
\Rightarrow e^- \gamma \gamma$) has been obtained in a similar fashion
\cite{morozov_elastic_1975, seipt_two-photon_2012, mackenroth_double_2013,
king_double_2015, dinu_double_2019}. As to $2\Rightarrow2$ reactions,
considerations existing in the literature are limited to very specific cases,
like a monochromatic or an almost monochromatic laser field, the weak-field
limit, a circular laser polarization, and/or so-called resonance processes (see,
e.g., Refs. \cite{oleinik_resonance_1967, oleinik_resonance_1968,
hartin_dissertation_2006, denisenko_interference_2006,
voroshilo_resonant_2016}).

\begin{figure}[t]
    \centering
    \raisebox{78pt}{\textbf{(a)}}
    \hspace{0.5cm}
    \includegraphics{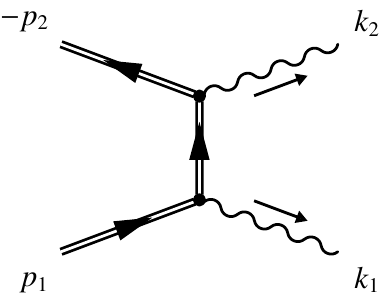}
    \hspace{2cm}
    \raisebox{78pt}{\textbf{(b)}}
    \hspace{0.5cm}
    \includegraphics{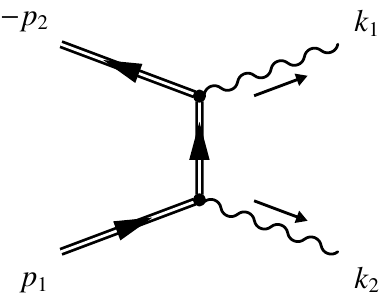}
    \caption{The leading-order Feynman diagrams for electron-positron
             annihilation into two photons in a plane-wave field: (a) the direct
             diagram, and (b) the exchange diagram. Double lines represent
             dressed (Volkov) wave functions and propagators (see the main text
             for details).}
    \label{fig:ep2g_diagrams}
\end{figure}

Here, we consider electron-positron annihilation into two photons, with the two
leading-order Feynman diagrams shown in \figref{fig:ep2g_diagrams}. We present
the first analytical results for a total cross section (in a sense explained
below) of $e^-e^+ \Rightarrow \gamma \gamma$ in a laser pulse represented as a
classical plane-wave (or null) field of a general shape. We provide an exact
expression for the contribution of the individual diagrams in
\figref{fig:ep2g_diagrams}, without taking into account the interference between
them. Keeping possible applications of our result to many-body-evolution
numerical codes in mind, we define the cross section in such a way that it has
the meaning of a local quantity. Furthermore, we use the example of $e^-e^+
\Rightarrow \gamma \gamma$ for establishing general features of the description
of second-order 2-to-2 collision processes in a plane-wave background field.

In contrast to nonlinear trident pair production and nonlinear double Compton
scattering, the reaction $e^-e^+ \Rightarrow \gamma \gamma$ does occur already
in vacuum. This may pose a technical problem, since the two parts (the vacuum
and field-dependent one) have different numbers of energy-momentum conservation
delta functions. Therefore, one might encounter a difficulty in dealing with the
different number of volume factors and in comparing and combining the two parts.
We show that it is possible to incorporate both into a single expression for the
total (local) cross section, which, in the limit of a vanishing external field,
reduces to the result, known in the literature for the vacuum case. Moreover,
unlike the mentioned second-order processes initiated by a single particle, for
$e^-e^+ \Rightarrow \gamma \gamma$ the intermediate fermion becomes real not in
one but in two different cases corresponding to the physical situations in which
either the electron or the positron first emits a final photon before
annihilating with the other particle into the second final photon. Using the
Schwinger proper time representation for the electron propagator, we express the
two-step and one-step contributions in a form, which has an advantage that it is
concise and involves only integrals with fixed limits. Another feature of 2-to-2
processes in a plane wave is the particular importance of taking into account
the fact of the localization of the incoming particles, which we carry out by
introducing normalized wave packets. The underlying reason is that the collision
of two particles in a plane wave is effectively a three-body collision and it is
important at which moment each participant arrives at the collision region and
if a collision region, as a microscopic region where all participants are for a
certain time and significantly interact, does exist at all.

This paper is organized as follows. In Sec.~\ref{sec:formalism} we introduce the
formalism. In Sec.~\ref{sec:cross_section} we consider the annihilation into two
photons of an electron and a positron, which are described by wave packets. We
find out the approximations that one needs to make in order to introduce a cross
section and provide a general expression for the cross section of the reaction
$e^-e^+\Rightarrow\gamma\gamma$. In Sec.~\ref{sec:step_split} the one- and
two-step contributions to the cross section are investigated. In
Sec.~\ref{sec:integrals} we elaborate on the evaluation of the integrals for the
process under consideration. The final result is presented in
Sec.~\ref{sec:result} and the limit of a vanishing background field is
considered in Sec.~\ref{sec:zero-field}. The discussion of the results and the
conclusions are presented in Sec.~\ref{sec:discussion} and
Sec.~\ref{sec:conclusions}, respectively. Five Appendixes contain explanations
of the notation and technical details.

Throughout the paper, Heaviside and natural units are used ($\epsilon_0 = \hbar
= c = 1$), $m$ and $e<0$ denote the electron mass and charge, respectively,
$\alpha = e^2/(4 \pi) \approx 1/137$ is the fine-structure constant.

\section{Formalism}
\label{sec:formalism}

The formalism, that we employ, combines light-cone quantization
\cite{bjorken_quantum_1971, mustaki_perturbative_1991, brodsky_quantum_1998,
burkardt_light-front_2002} and Furry picture \cite{furry_bound_1951,
fradkin_unstable_1991} (a detailed discussion of the formalism utilized here is
provided in Ref. \cite{bragin_dissertation_2019}). With the quantization on the
light cone, a plane-wave background and particularly momentum conservation laws
are naturally included into the calculations (see Refs. \cite{dinu_trident_2018,
dinu_double_2019} for an application of light-cone quantization to trident and
double Compton scattering). Also, the light-cone representation of the
electromagnetic interaction via three types of vertices (see Appendix
\ref{sec:light-cone_quantization}), or, equivalently, the representation of the
electron propagator (and also of the photon one) as combination of
noninstantaneous and instantaneous terms (this can be done within the
instant-form quantization as well \cite{kogut_quantum_1970,
mantovani_revisiting_2016}; see also Refs. \cite{seipt_two-photon_2012,
mackenroth_double_2013, hartin_fierz_2016}) allows one to write the spinor
prefactors via fermion dressed momenta (see below), and, as a consequence, the
final expressions formally have no explicit dependence on the background field
and asymptotic fermion momenta. In this respect, the obtained result is similar
to the ones usually derived in vacuum, where the final expressions depend on the
particle four-momenta in the form of Mandelstam variables
\cite{berestetskii_qed_1982}.

The laser field is described classically by the field tensor $F^{\mu\nu}(\phi)$,
which is a function of the scalar product $\phi = k_0x$, with $k_0^\mu$ being
the characteristic wave four-vector of the field or, in the quantum language,
the characteristic four-momentum of a laser photon ($k_0^2 = k_0^\mu k_{0\mu} =
0$) and $x^\mu$ being a position four-vector. We assume that $F^{\mu\nu}(\phi)$
does not contain a zero-frequency contribution, i.e., the integral of
$F^{\mu\nu}(\phi)$ over the whole real axis vanishes. Then the most general form
of $F^{\mu\nu}(\phi)$ is given by
\begin{equation}
    F^{\mu\nu}(\phi) = \sum_{i=1,2} f_i^{\mu\nu} \psi_i'(\phi),
\end{equation}
where $f_i^{\mu \nu} = k_0^\mu a_i^\nu - k_0^\nu a_i^\mu$, the four-vectors
$a_i^\mu$ define the amplitude of the field in two polarization directions ($k_0
a_i = 0$, $a_1 a_2 = 0$), and the functions $\psi_i'(\phi) = \dd
\psi_i(\phi)/\dd \phi$ characterize its shape [$|\psi_i'(\phi)| \lesssim 1$]. In
the following, each of the indices $i,j$ always take the values 1, 2.

For an arbitrary four-vector $a^\mu$, we define light-cone coordinates as $a^+ =
a \eta$, $a^- = a \bar{\eta}$, and $a^i = - ae_i$ [$\vec{a}^\perp = (a^1,a^2)$],
where $\{\eta^\mu, \bar{\eta}^\mu, e_1^\mu, e_2^\mu\}$ is a light-cone basis
(see Appendix \ref{sec:light-cone_quantization} for details). Below, we employ
the canonical light-cone basis, which is given by
\cite{meuren_dissertation_2015}
\begin{equation}
    \label{eq:canonical_light-cone_basis}
    \eta^\mu = \frac{k_0^\mu}{m},
    \quad
    \bar{\eta}^\mu = \frac{P^\mu}{P^+} - \frac{P^2\eta^\mu}{2 P^{+2}},
    \quad
    e_1^\mu = \frac{P_\lambda f_1^{\lambda \mu}}{m P^+ \sqrt{-a_1^2}},
    \quad
    e_2^\mu = \frac{P_\lambda f_2^{\lambda \mu}}{m P^+ \sqrt{-a_2^2}},
\end{equation}
where the four-vector $P^\mu$ is arbitrary except that $P^+ \neq 0$. The
calculations are greatly simplified if one chooses
\begin{equation}
    \label{eq:qmu_specified}
    P^\mu = p_1^\mu + p_2^\mu,
\end{equation}
which implies $\vec{p}_2^\perp + \vec{p}_1^\perp = \vec{k}_2^\perp +
\vec{k}_1^\perp = \vec{0}$ [see \eqref{eq:canonical_light-cone_basis}
and Appendix~\ref{sec:light-cone_quantization}; also note that $f_i^{\mu \nu}$
is an antisymmetric tensor]. Here, $p_1^\mu$ and $p_2^\mu$ are the asymptotic
four-momenta of the incoming electron and positron outside the plane wave,
respectively, whereas $k_1^\mu$ and $k_2^\mu$ are the four-momenta of the final
photons (see \figref{fig:ep2g_diagrams} and note that in the following we employ
wave packets for the electron and positron, and therefore $p_1^\mu$ and
$p_2^\mu$ will be ultimately identified with the central four-momenta).

Since $\eta^\mu = k_0^\mu/m$, the laser phase is $\phi = mx^+$ and the field
$F^{\mu \nu}(\phi)$ depends only on the light-cone time. With the adoption of
the light-cone gauge $A^+(x) = 0$, the four-vector potential for
$F^{\mu\nu}(\phi)$ reads
\begin{equation}
    \label{eq:potential}
    A^\mu(\phi) = \sum_i a_i^\mu \psi_i(\phi).
\end{equation}
In the following, we assume $A^\mu(-\infty)=0$, which implies
$\psi_i(-\infty)=0$ [together with the fact of the absence of a zero-frequency
contribution in $F^{\mu\nu}(\phi)$ this implies that also $A^\mu(\infty)=0$, and
therefore $\psi_i(\infty)=0$].

The solution of the Dirac equation with the classical field (\ref{eq:potential})
is the Volkov solution \cite{volkov_1935}. We write the positive-energy one in
the form \cite{hartin_fierz_2016}
\begin{equation}
    \label{eq:volkov_solution_positive}
    \psi_{p \sigma}(x) = K_p(\phi) \frac{u_{p \sigma}}{\sqrt{2p^+}}
    \ee^{iS_p(x)},
\end{equation}
with
\begin{equation}
    K_p(\phi) = [\gamma \pi_p(\phi) + m] \frac{\gamma^+}{2p^+}, 
    \quad
    S_p(x)=-px-\mcS_p(\phi), 
    \quad
    \mcS_p(\phi) 
    = \int\displaylimits_{-\infty}^{\phi} \dd \beta\, 
    \left( \frac{epA(\beta)}{mp^+} - \frac{e^2A^2(\beta)}{2mp^+} \right),
\end{equation}
and the negative-energy one in an analogous way (see Appendix
\ref{sec:light-cone_quantization}). Note that the phase $S_p(x)$ is the
classical action of an electron in the plane wave and that the dressed
four-momentum $\pi_p^\mu(\phi) = -\partial^{\mu}S_p(x)-eA^{\mu}(\phi)$ is the
corresponding solution of the Lorentz equation. It is given by
\begin{equation}
    \label{eq:dressed_momentum}
    \pi_p^\mu(\phi) = p^\mu - eA^\mu(\phi) 
    + \eta^\mu\left(\frac{e pA(\phi)}{p^+} - \frac{e^2 A^2(\phi)}{2p^+}\right), 
\end{equation}
such that $\pi_p^2(\phi)=p^2$, $\pi_p^+(\phi)=p^+$. The free Dirac bispinor
$u_{p \sigma}$ is normalized such that $\bar{u}_{p\sigma}u_{p \sigma'} =
2m\delta_{\sigma\sigma'}$, $\bar{u}_{p\sigma}\gamma^\mu u_{p\sigma'} =
2p^\mu\delta_{\sigma\sigma'}$, $\sum_\sigma u_{p\sigma}\bar{u}_{p\sigma} =
\gamma p + m$, where $\bar{u}_{p \sigma} = u_{p \sigma}^\dagger \gamma^0$ and
the dagger denotes the Hermitian conjugate, and analogous expressions are valid
for the negative-energy bispinor $v_{p \sigma}$ \cite{berestetskii_qed_1982}.

The fermion field $\psi(x)$ is expanded in the basis set of the Volkov wave
functions (\ref{eq:volkov_solution_positive}) (and analogous ones for
negative-energy states) and, as a consequence, in all diagrams free fermion
lines are replaced with the corresponding Volkov ones \cite{furry_bound_1951,
fradkin_unstable_1991} (details on the quantization are given in Appendix
\ref{sec:light-cone_quantization}).

Though in electrodynamics, quantized on the light cone, there are three types of
vertices, for our purposes it is convenient to combine them in the form of
propagators. Then we have only the usual three-point QED vertex, but each
electron and photon Feynman propagator consists of two terms
\cite{kogut_quantum_1970, mantovani_revisiting_2016} (see also Refs.
\cite{seipt_two-photon_2012, mackenroth_double_2013, hartin_fierz_2016}), in
particular, for the electron propagator $G(x_2,x_1)$ we have $G(x_2,x_1) =
G^{\text{(ni)}}(x_2,x_1) + G^{\text{(in)}}(x_2,x_1)$, with
$G^{\text{(ni)}}(x_2,x_1)$ being a noninstantaneous (propagating) part,
\begin{equation}
    \label{eq:electron_propagator_ni}
    G^{(\text{ni})}(x_2, x_1) = \int \frac{\dd^4 p}{(2 \pi)^4} 
    \ee^{-ip(x_2 - x_1) -i\mcS_p(\phi_2,\phi_1)} K_p(\phi_2) 
    \frac{\gamma \tilde{p} + m}{p^2 - m^2 + i \epsilon} \bar{K}_p(\phi_1),
\end{equation}
where $\bar{K}_p(\phi)=\gamma^0K_p^\dagger(\phi)\gamma^0$,
and with $G^{\text{(in)}}(x_2,x_1)$ being an instantaneous part,
\begin{equation}
    G^{(\text{in})}(x_2, x_1) = \int \frac{\dd^4 p}{(2 \pi)^4} 
    \ee^{-ip(x_2 - x_1) -i\mcS_p(\phi_2,\phi_1)}
    \frac{\gamma^+}{2p^+}.
\end{equation}
Here, $\mcS_p(\phi_2,\phi_1) = \mcS_p(\phi_2) -\mcS_p(\phi_1)$ and
\begin{equation}
    \tilde{p}^\mu = \left( p^+, \frac{\boldsymbol{p}^{\perp2} + m^2}{2p^+},
    \boldsymbol{p}^\perp \right),
\end{equation}
such that $\tilde{p}^2 = m^2$.

Below, we will employ the classical intensity parameters
\cite{ritus_quantum_effects_1985, di_piazza_review_2012}
\begin{equation}
    \label{eq:xi_defined}
    \xi_i = \frac{|e|\sqrt{-a_i^2}}{m},
\end{equation}
and we also introduce $\xi = \sqrt{\xi_1^2 + \xi_2^2}$. Other parameters
characterizing the scattering process are the quantum nonlinearity parameters,
which are defined as $\chi_{i}=p_i^+\xi/m$ for the fermions, and analogously for
the photons \cite{ritus_quantum_effects_1985, di_piazza_review_2012}. Note that
by considering the interaction with the quantized photon field to leading order,
we implicitly assume that the quantum nonlinearity parameters are much smaller
than $1/\alpha^{3/2}\approx 1600$, such that this interaction can be treated
perturbatively. This assumption is reasonable for current and near-future
laser-based setups (for discussions of the fully nonperturbative regime, see,
e.g., Refs.~\cite{fedotov_conjecture_2017, yakimenko_prospect_2019,
podszus_high-energy_2019, ilderton_note_2019, baumann_probing_2019,
di_piazza_testing_2020}).

For a process with two incoming particles, the classical intensity parameters
and the quantum nonlinearity parameters do not exhaust the list of quantities,
that are necessary for describing the scattering (even when considering an
observable obtained by averaging/summing over the discrete quantum numbers and
by integrating over the final momenta). We introduce the additional parameters
$t_i(\phi)$, which are given by \cite{bragin_dissertation_2019}
\begin{equation}
    \label{eq:tparameter_defined}
    t_i(\phi)
    = \frac{|e| \pi_{e,p_1}^\mu(\phi) f_{i\mu \nu} \pi_{-e,p_2}^\nu(\phi)}
    {\xi_i m^3 (p_1^+ + p_2^+)},
\end{equation}
where $\pi_{e,p_1}^\mu(\phi) = \pi_{p_1}^\mu(\phi)$ and $\pi_{-e,p_2}^\mu(\phi)
= -\pi_{-p_2}^\mu(\phi)$ are the dressed four-momenta of the electron and the
positron, respectively. The asymptotic values of $t_i(\phi)$ are denoted as
$t_i$, they have been employed in the literature
before~\cite{ritus_quantum_effects_1985}.

The parameters $t_i(\phi)$ have a particularly clear physical interpretation if
we use the canonical light-cone basis (\ref{eq:canonical_light-cone_basis}) with
$P^\mu$ from \eqref{eq:qmu_specified}. With this choice, we have
$\vec{\pi}_{e,p_1}^\perp(\phi) + \vec{\pi}_{-e,p_2}^\perp(\phi) =
\vec{p}_1^\perp + \vec{p}_2^\perp = \vec{0}$ and $t_i(\phi) =
\pi_{p_1}^i(\phi)/m$, i.e., $t_1(\phi)$ and $t_2(\phi)$ correspond to the
transverse dressed momentum components of the incoming particles (with respect
to the laser-pulse propagation direction).

\section{Cross section}
\label{sec:cross_section}

As has been mentioned in the Introduction, the result of a collision of an
electron, a positron, and a finite-duration laser pulse depends on the existence
of a collision region and the time of arrival of each participant at this
region. Thus, in the most general setup, one cannot rely on the description of
the incoming particles via monochromatic plane waves, since they have an
infinite temporal and spatial extent.

Therefore, in order to consistently describe the reaction $e^-e^+ \Rightarrow
\gamma \gamma$, we represent the electron and the positron as normalized wave
packets with central on-shell four-momenta $p_1^\mu$ and $p_2^\mu$,
respectively. A positive-energy wave packet $\Psi_p(x)$ with the central
four-momentum $p^\mu$ is constructed according to
\begin{equation}
    \label{eq:wave_packet_definition}
    \Psi_p(x) = \int \frac{\tilde{\dd}^3 q}{(2 \pi)^3} 
    \tilde{f}_p(q) \psi_q(x),
\end{equation}
where $\tilde{f}_p(q)$ is the momentum distribution density and $\psi_q(x)$ is
the positive-energy Volkov state (\ref{eq:volkov_solution_positive}) with
four-momentum $q^\mu$ (for the definition of $\tilde{\dd}^3 q$ see Appendix
\ref{sec:light-cone_quantization}). Note that Volkov states are on-shell such
that $q^- = (\vec{q}^{\perp2}+m^2)/(2q^+)$, i.e., $\tilde{f}_p(q)$ depends on
$\vec{q}^\perp$ and $q^+$ only, but for simplicity, we write $\tilde{f}_p(q)$ as
a function of $q^{\mu}$. The fact that the function $\tilde{f}_p(q)$ is centered
around the on-shell four-momentum $p^{\mu}$ has to be intended analogously.
Correspondingly, one can also define negative-energy wave packets. We refer to
Appendix \ref{sec:wave_packets} for further details about the general properties
of the wave packets $\Psi_p(x)$.

The polarization degrees of freedom of both incoming (outgoing) particles are
averaged (summed) in the final expressions, with the assumption of the initial
states being unpolarized, and therefore, for notational brevity, we suppress the
subscripts for these degrees of freedom.

As already mentioned, the final photon four-momenta are $k_1^\mu$ and $k_2^\mu$
($k^2_1=k^2_2=0$). The $S$-matrix element corresponding to the diagrams in
\figref{fig:ep2g_diagrams} can be written as
\begin{equation}
    \label{eq:sfi_initial}
    S_{fi} = i \int
    \frac{\tilde{\dd}^3 q_2}{(2 \pi)^3} \frac{\tilde{\dd}^3 q_1}{(2 \pi)^3}
    \tilde{f}_2^*(q_2) \tilde{f}_1(q_1)
    \int \dd^4 x_2 \dd^4 x_1 \, \tilde{T}(x_2,x_1,q_2,q_1),
\end{equation}
where we have introduced the shorthand notation $\tilde{f}_1(q) =
\tilde{f}_{p_1}(q)$ and $\tilde{f}_2^*(q) = \tilde{f'}_{p_2}^*(q)$ for the
electron and positron wave-packet momentum distributions $\tilde{f}_{p_1}(q)$
and $\tilde{f'}_{p_2}^*(q)$, respectively (an asterisk indicates the complex
conjugate), and
\begin{align}
    \label{eq:tfi}
    \tilde{T}(x_2,x_1,q_2,q_1) 
    = \int \frac{\dd^4 p_3}{(2 \pi)^4} 
    \frac{M^{\text{direct}}(\phi_2,\phi_1,q_2,q_1)}{4\sqrt{k_2^+k_1^+q_2^+q_1^+}}
    \exp[&i(k_2 - p_3 - q_2)x_2 + i(k_1 + p_3 - q_1) x_1
    \nonumber\\&
    - i\mcS_{p_3}(\phi_2,\phi_1) - i\mcS_{q_1}(\phi_1) 
    + i\mcS_{-q_2}(\phi_2)]
    \nonumber\\&+
    \{\gamma_1 \leftrightarrow \gamma_2\},
\end{align}
with
\begin{align}
    \label{eq:tfi_direct_prefactor}
    M^{\text{direct}}(\phi_2,\phi_1,q_2,q_1) = &-e^2
    \bar{v}_{q_2} \left[ K_{-q_2p_3}^\mu(\phi_2) 
    \frac{\gamma \tilde{p}_3 + m}{p_3^2-m^2+i \epsilon} K_{p_3q_1}^\nu(\phi_1)
    \right] u_{q_1} \epsilon_{2 \mu}^* \epsilon_{1 \nu}^*
    \nonumber\\
    &-e^2 \bar{v}_{q_2} \left[
    \frac{K_{-q_2q_1}^{\mu \nu}(\phi_2,\phi_1)}{2p_3^+} \right] u_{q_1}
    \epsilon_{2 \mu}^* \epsilon_{1 \nu}^*.
\end{align}
Here and below, $\phi_i = k_0 x_i = mx_i^+$ and the term $\{\gamma_1
\leftrightarrow \gamma_2\}$ corresponds to the exchange diagram with the photon
quantum numbers swapped (see \figref[b]{fig:ep2g_diagrams}). Also, the functions
$K_{pp'}^\mu(\phi)$ and $K_{pp'}^{\mu \nu}(\phi,\phi')$ are given by
\begin{equation}
    K_{pp'}^\mu(\phi) = \bar{K}_{p}(\phi) \gamma^\mu K_{p'}(\phi),
    \quad
    K_{pp'}^{\mu \nu}(\phi,\phi') 
    = \bar{K}_{p}(\phi) \gamma^\mu \gamma^+ \gamma^\nu K_{p'}(\phi').
\end{equation}

In the following and analogously to the vacuum case (see, e.g.,
Refs.~\cite{goldberger_watson_1964, itzykson_zuber_1980}), we assume the
momentum distributions of the electron and the positron being sufficiently
narrowly peaked around the central four-momenta and the detectors not being
sensitive enough to resolve the final momenta within the widths of such
distributions, such that we can in particular replace the four-momenta $q_i^\mu$
with the central ones in relatively slowly varying functions, i.e.,
\begin{equation}
    M^{\text{direct}}(\phi_2,\phi_1,q_2,q_1)/\sqrt{q_2^+q_1^+} \approx
    M^{\text{direct}}(\phi_2,\phi_1)/\sqrt{p_2^+p_1^+},
\end{equation}
where $M^{\text{direct}}(\phi_2,\phi_1) =
M^{\text{direct}}(\phi_2,\phi_1,p_2,p_1)$, and we do the same for the exchange
term as well.

The total probability, obtained as the modulus squared of
\eqref{eq:sfi_initial}, averaged over the initial polarization states and summed
over all final polarization and momentum states, can be written as
\begin{align}
    \label{eq:w_total_general}
    W &\approx \frac14 \sum_{\text{qn}} \left|\int \dd^4 x_2 \dd^4 x_1 
    F_2^*(x_2) F_1(x_1) \tilde{T}(x_2,x_1,p_2,p_1)\right|^2
    \nonumber\\&=
    \frac14 \sum_{\text{qn}} \int \dd^4 x_2 \dd^4 x_1 \dd^4 x'_2 \dd^4 x'_1
    F_2^*(x_2) F_2(x'_2) F_1(x_1) F_1^*(x'_1) 
    \tilde{T}(x_2,x_1,p_2,p_1) \tilde{T}^*(x'_2,x'_1,p_2,p_1),
\end{align}
where the abbreviation ``qn'' indicates that the sum/integral is taken over the
discrete quantum numbers of the initial and final particles and the momenta of
the final photons. Also, in \eqref{eq:w_total_general} we have introduced the
electron and positron wave-packet amplitudes $F_1(x_1)$ and $F_2^*(x_2)$ in
configuration space, which are defined analogously to the vacuum case
\cite{goldberger_watson_1964}, e.g., for an electron we have
\begin{equation}
    \label{eq:wave_packet_amplitude}
    F_p(x) = \int \frac{\tilde{\dd}^3 q}{(2 \pi)^3} \tilde{f}_p(q) 
    \exp[-i(q-p)x - i\mcS_q(\phi) + i\mcS_p(\phi)]
\end{equation}
for a given momentum distribution $\tilde{f}_p(q)$. The scalar wave packet in
configuration space is given by
\begin{equation}
    \label{eq:wave_packet_position}
    f_p(x) = \int \frac{\tilde{\dd}^3 q}{(2 \pi)^3} \tilde{f}_p(q) 
    \exp[-iqx - i\mcS_q(\phi)] = F_p(x) \exp[-ipx - i\mcS_p(\phi)]
\end{equation}
(for a positron, the expressions are analogous). Note that $|f_p(x)|^2 =
|F_p(x)|^2$ is the (time-dependent) particle density. The properties of the
particle density $|F_p(x)|^2$ are discussed in Appendix \ref{sec:wave_packets},
and we only recall here that for a narrow wave packet, under the condition that
also $|f_p(x)|^2$ is sufficiently peaked in configuration space, the center of
the distribution $|f_p(x)|^2$ follows the classical trajectory of an electron in
a given plane wave (see Appendix \ref{sec:wave_packets} for further details).

In principle, \eqref{eq:w_total_general} is the expression one needs to employ
in order to evaluate the total probability of the process under consideration.
However, depending on the widths of the wave packets and on the formation
lengths of the integrals in the space-time variables, one can achieve further
simplifications.

The first step is to assume that the wave packets are sufficiently narrow (in
momentum space), that on the formation length of a single-vertex process
(essentially, a process obtained by cutting the propagator line, see
\figref{fig:ep2g_diagrams}) one can neglect the interference among the wave
packets, i.e.,
\begin{equation}
    \label{eq:wave_packet_approximation1}
    F_1(x_1) F_1^*(x'_1) = F_1(X_1-\delta_1/2) F_1^*(X_1+\delta_1/2) 
    \approx |f_1(X_1)|^2
\end{equation}
and analogously for the positron wave-packet amplitudes, where
\begin{equation}
    X_i^\mu = (x_i^\mu + x_i'^\mu)/2,
    \quad
    \delta_1^\mu = {x_1'}^\mu - x_1^\mu,
    \quad
    \delta_2^\mu = x_2^\mu - {x_2'}^\mu.
\end{equation}
Note that the approximation (\ref{eq:wave_packet_approximation1}) is not assumed
to be valid for all values of $\delta_1^\mu$. It is assumed to be valid for
$\delta_1^\mu$ only within the formation region of the integral in this
variable, i.e., within an effective part of the whole space which mostly
contributes to the value of the integral.

We also point out that the assertion in \eqref{eq:wave_packet_approximation1}
[and the corresponding one for $F_2(x_2')F_2^*(x_2)$] is a more complicated
statement than in vacuum, in the sense that the typical scale of $\delta_1^\mu$
(and of $\delta_2^\mu$ for the positron) depends in general on the form and on
the intensity of a considered background field, and Eq.
(\ref{eq:wave_packet_approximation1}) results from an interplay between the
scale introduced by the field and the scale of the wave packets (details are
given in Appendix \ref{sec:conditions}).

Under the approximation (\ref{eq:wave_packet_approximation1}) and an analogous
one for the positron, the total probability (\ref{eq:w_total_general}) reads
\begin{equation}
    \label{eq:w_total_two-point}
    W \approx \int \dd^4 X_2 \dd^4 X_1 \, |f_2(X_2)|^2 |f_1(X_1)|^2 W(X_2,X_1),
\end{equation}
with the two-point probability distribution
\begin{equation}
    \label{eq:w_two-point}
    W(X_2,X_1) = \frac14 \sum_{\text{qn}}
    \int \dd^4 \delta_2 \dd^4 \delta_1 \,
    \tilde{T}(x_2,x_1,p_2,p_1) \tilde{T}^*(x'_2,x'_1,p_2,p_1).
\end{equation}

An additional simplification is attained under the assumption, that on a typical
distance between $X_1^\mu$ and $X_2^\mu$ (in essence, on the typical distance
between the two single-vertex processes, see \figref{fig:ep2g_diagrams}) the
wave packets do not change significantly, i.e.,
\begin{equation}
    \label{eq:wave_packets_one-point}
    |f_2(X_2)|^2 |f_1(X_1)|^2 
    = |f_2(x+\delta/2)|^2 |f_1(x-\delta/2)|^2 
    \approx |f_2(x)|^2 |f_1(x)|^2,
\end{equation}
where
\begin{equation}
    \label{eq:x_delta_defined}
    x^\mu = (X_2^\mu + X_1^\mu)/2,
    \quad
    \delta^\mu = X_2^\mu - X_1^\mu.
\end{equation}
Then \eqref{eq:w_total_two-point} transforms into
\begin{equation}
    \label{eq:w_total_one-point}
    W \approx \int \dd^4 x \, |f_2(x)|^2 |f_1(x)|^2 W(x),
\end{equation}
where
\begin{equation}
    \label{eq:w_local}
    W(x) = W(\phi) = \frac14 \sum_{\text{qn}}
    \int \dd^4 \delta \dd^4 \delta_2 \dd^4 \delta_1 \,
    \tilde{T}(x_2,x_1,p_2,p_1) \tilde{T}^*(x'_2,x'_1,p_2,p_1).
\end{equation}
Equation (\ref{eq:w_total_one-point}) is the approximation that is commonly used
for the description of scattering in vacuum and that allows us to define a cross
section, a quantity, which characterizes the process itself without relying on
the precise shape of the wave packets \cite{goldberger_watson_1964,
itzykson_zuber_1980}. We stress that in a background field the assumption
(\ref{eq:wave_packets_one-point}) can be restrictive as the intermediate
particle may become real, and hence $\delta^\mu$ can have a macroscopic scale,
i.e., of the order of the extension of the background field. For the highly
nonlinear regime ($\xi \gg 1$), semiquantitative estimations imply (see Appendix
\ref{sec:conditions} for details) that if one excludes the contribution of the
case of the intermediate particle being real, both approximations
(\ref{eq:wave_packet_approximation1}) and (\ref{eq:wave_packets_one-point}) are
valid as soon as the relations $|\Delta \vec{p}_1^\perp| \ll 
\max(m, |\vec{\pi}_{p_1}^{\perp}(\phi)|)$, $\Delta p_1^+ \ll p_1^+$ for the
electron wave packet and analogous ones for the positron wave packet are
fulfilled ($\Delta p_1^+$ and $\Delta \vec{p}_1^\perp$ are the corresponding
wave-packet widths).

Now, it is worth pointing out an additional difference with the vacuum case. In
the latter case, in fact, the quantity $W(x)$ is independent of the coordinates
and therefore non-negative \cite{goldberger_watson_1964, itzykson_zuber_1980}.
In contrast to this, the quantity $W(\phi)$ here explicitly depends on the
light-cone time (via $\phi$) and it can be negative for some values of $\phi$.
Thus, generally speaking, the quantity
\begin{equation}
    w(x) = |f_2(x)|^2 |f_1(x)|^2 W(\phi)
\end{equation}
cannot be interpreted as a probability per unit time and unit volume. However,
it can be seen as a quantity, which generalizes this probability and which
entails interference effects among contributions from different points of the
particles trajectory in the plane wave, and therefore may become negative. This
is somewhat similar to the relation between a classical phase-space distribution
and the Wigner distribution, with the latter generalizing the former and,
indeed, being also potentially negative \cite{wigner_quantum_1932}.

Furthermore, we can define a generalized (local) cross section, which, though
not being directly an observable quantity, since it can become negative, is a
useful theoretical tool for investigating the influence of the external field on
the scattering process. We follow the approach in the instant-form quantization
in vacuum, where the cross section is obtained from the probability per unit
time and unit volume by dividing it by the factor
$|g_2(x)|^2|g_1(x)|^2I/(p_2^0p_1^0)$, where $I = \sqrt{(p_2p_1)^2-m^4}$ and
$g_i(x)$ are wave packets in the instant form \cite{itzykson_zuber_1980,
berestetskii_qed_1982}. Then in our case we can analogously introduce the local
cross section as
\begin{equation}
    \label{eq:cross_section}
    \sigma(\phi) 
    = \frac{p_2^+p_1^+}{|f_2(x)|^2 |f_1(x)|^2 I(\phi)} w(x)
    = \frac{p_2^+p_1^+}{I(\phi)} W(\phi),
\end{equation}
where the invariant $I(\phi)$ reads
\begin{equation}
    I(\phi) = \sqrt{[\pi_{-e,p_2}(\phi) \pi_{e,p_1}(\phi)]^2 - m^4}.
\end{equation}
Below, we explicitly verify (except for the interference term, as has been
pointed out in the Introduction) that in the absence of the background field the
cross section (\ref{eq:cross_section}) reduces to the one, known for the vacuum
case in the instant-form quantization. One should also keep in mind that the
choice of the invariant $I(\phi)$ implies that the cross section is normalized
to the flux coming into the point $x^\mu$ inside the laser field and in this
sense is a local quantity. This can be useful, for instance, in the analysis of
the importance of the studied process in the development of QED cascades where
the colliding particles are produced inside the field. However, if one would
like to consider a beam-beam collision experiment in the presence of a laser
field, then the use of the vacuum counterpart $I$ in place of $I(\phi)$ could be
more convenient. The total probability $W$ is of course independent of this
choice.

We emphasize that Eqs. (\ref{eq:w_total_two-point}) and
(\ref{eq:w_total_one-point}) are not ensured to provide a positive result in a
general case, i.e., without taking into account the validity of the
approximations (\ref{eq:wave_packet_approximation1}) and
(\ref{eq:wave_packets_one-point}), and that one has to ultimately rely on the
probability $W$ in \eqref{eq:w_total_general}, if these approximations break
down.

With the use of Eqs. (\ref{eq:tfi}) and (\ref{eq:tfi_direct_prefactor}), we
obtain for the cross section:
\begin{equation}
    \label{eq:cross_section_final}
    \sigma(\phi) = \int\displaylimits_0^{p_2^+ + p_1^+} 
    \frac{\dd k_1^+}{2 \pi} \int \frac{\dd^2 k_1^\perp}{(2 \pi)^2} 
    \frac{1}{32k_2^+k_1^+I(\phi)}
    \int \dd \delta^+ \dd \delta_2^+ \dd \delta_1^+ \,
    \frac14 \sum\displaylimits_{\sigma_i, \lambda_i} 
    \tilde{M}(\phi_2,\phi_1,p_2,p_1)\tilde{M}^*(\phi_2',\phi_1',p_2,p_1),
\end{equation}
where $k_2^{(+,\perp)} = (p_2+p_1-k_1)^{(+,\perp)}$, $\sigma_i$ and $\lambda_i$
denote the polarization states of the incoming and outgoing particles,
respectively, and we have divided the result by 2, in order to compensate for
the double counting of the final states of the two identical particles. The
reduced matrix element $\tilde{M}(\phi_2,\phi_1,p_2,p_1)$ is given by
\begin{align}
    \label{eq:mfi_tilde}
    \tilde{M}(\phi_2,\phi_1,p_2,p_1) 
    = \int \frac{\dd p_3^-}{2 \pi} M^{\text{direct}}(\phi_2,\phi_1)
    \exp[&i(k_2^- - p_3^- - p_2^-)x_2^+ + i(k_1^- + p_3^- - p_1^-) x_1^+
    \nonumber\\&
    - i\mcS_{p_3}(\phi_2,\phi_1) - i\mcS_{p_1}(\phi_1) + i\mcS_{-p_2}(\phi_2)]
    \nonumber\\&+
    \{\gamma_1 \leftrightarrow \gamma_2\},
\end{align}
where $p_3^{(+,\perp)} = (p_1-k_1)^{(+,\perp)} = (k_2-p_2)^{(+,\perp)}$.

The quantity $\tilde{M}(\phi_2,\phi_1,p_2,p_1)$ in \eqref{eq:mfi_tilde} contains
four distinct terms because $M^{\text{direct}}(\phi_2,\phi_1)$ alone consists of
a noninstantaneous and an instantaneous contributions, corresponding to the
first and to the second term on the right-hand side of
\eqref{eq:tfi_direct_prefactor}, respectively. Taking the modulus squared yields
16 terms. However, only eight of them are different after we sum over the states
of the final photons, i.e.,
\begin{equation}
    \label{eq:sigma_terms}
    \sigma(\phi) = \sigma^{\text{dd}}(\phi) + \sigma^{\text{ee}}(\phi)
    + \sigma^{\text{de}}(\phi) + \sigma^{\text{ed}}(\phi)
    = 2\sigma^{\text{dd}}(\phi) + 2\sigma^{\text{de}}(\phi),
\end{equation}
where $\sigma^{\text{dd}}(\phi)$ is the contribution, arising from squaring the
amplitude for the direct diagram (see \figref[a]{fig:ep2g_diagrams}), and can be
written as
\begin{equation}
    \label{eq:sigma_dd_terms}
    \sigma^{\text{dd}}(\phi) 
    = \sigma^{\text{nndd}}(\phi) + \sigma^{\text{nidd}}(\phi)
    + \sigma^{\text{indd}}(\phi) + \sigma^{\text{iidd}}(\phi).
\end{equation}
The four contributions in \eqref{eq:sigma_dd_terms} are obtained via squaring
corresponding parts of the amplitude [$\sigma^{\text{nndd}}(\phi)$ originates
from squaring the noninstantaneous direct term, $\sigma^{\text{nidd}}(\phi)$
from the product of the noninstantaneous and complex-conjugate instantaneous
direct terms, etc.], with subsequent rearrangements, as described below and in
Appendix~\ref{sec:traces}. The other contributions in \eqref{eq:sigma_terms} can
be written down analogously. In the following, we only consider
$\sigma^{\text{dd}}(\phi)$. We note that for the differential quantities the
interference terms ``de'' and ``ed'' lead to an enhancement of the cross section
by a factor of 2 in the case of the final photons being in the same state. On
the other hand, at least in an ultrarelativistic setup, the available phase
space is typically so large that one might expect that the integrated
interference term $\sigma^{\text{de}}(\phi)$ should give a negligible
contribution. Indeed, e.g., in the vacuum case, the interference contribution
for the total cross section is relatively large only for mildly relativistic
collisions \cite{berestetskii_qed_1982}. If we assume a similar behavior in our
case, then we should expect that the term $\sigma^{\text{de}}(\phi)$ might be
nonnegligible only for some $\sqrt{s(\phi)} \sim m$, where the invariant mass
squared $s(\phi)$ in the field is defined as
\begin{equation}
    \label{eq:inv_mass_sqr_field}
    s(\phi) = [\pi_{-e,p_2}(\phi) + \pi_{e,p_1}(\phi)]^2 
    = \frac{m^2(p_2^+ + p_1^+)^2}{p_2^+p_1^+}[1+\vec{t}^{\perp2}(\phi)],
\end{equation}
with $\vec{t}^{\perp2}(\phi) = t_1^2(\phi)+t_2^2(\phi)$. It follows that
if $p_1^+ \sim p_2^+$ and $\vec{t}^{\perp2}(\phi) \lesssim 1$, the interference
term might provide a somewhat sizable contribution. However, for the phase
average $\langle t_i^2(\phi) \rangle$ we have $\langle t_i^2(\phi) \rangle
\approx t_i^2 + \xi_i^2 \langle \psi_i^2(\phi) \rangle \gg 1$ if $\xi_i \gg 1$
[we assume that $\langle \psi_i(\phi) \rangle \ll 1$]. This implies that in the
highly nonlinear regime, i.e., in the regime of $\xi \gg 1$, and for
sufficiently long laser pulses, common values of $|\vec{t}^\perp(\phi)|$ are
much larger than unity. Therefore, if one considers dynamics over several laser
periods, one might expect that on average the term $\sigma^{\text{de}}(\phi)$
can be neglected.

Summing over the final photon polarizations results in the replacement
\begin{equation}
    \label{eq:replacement_photon_sum}
    \epsilon^\mu_{i} \epsilon^{*\nu}_{i} \to - g^{\mu \nu}
\end{equation}
(we discard the terms proportional to $k_i^\mu$ and $k_i^\nu$ due to the Ward
identity).

Averaging over the polarization states of the initial particles results in the
replacements~\cite{berestetskii_qed_1982}
\begin{equation}
    \label{eq:replacement_fermion_average}
    u_{p_1} \bar{u}_{p_1} \to \rho_{p_1}, 
    \quad 
    v_{p_2} \bar{v}_{p_2} \to \rho^{(-)}_{p_2},
\end{equation}
and taking the trace over the bispinor part of $\tilde{M}(\phi_2,\phi_1,p_2,p_1)
\tilde{M}^*(\phi_2',\phi_1',p_2,p_1)$. The quantities $\rho_{p_1}$ and
$\rho^{(-)}_{p_2}$ denote the electron and positron density matrices,
respectively. In the case of the initial particles being unpolarized, we have
\begin{equation}
    \rho_{p_1} = \frac{1}{2}(\gamma p_1 + m), 
    \quad 
    \rho^{(-)}_{p_2} = - \rho_{-p_2} = \frac{1}{2}(\gamma p_2 - m).
\end{equation}
Upon squaring the noninstantaneous part of the direct diagram, we obtain
\begin{equation}
    \label{eq:mm_nndd}
    \frac14 \sum_{\sigma_i, \lambda_i} 
    \tilde{M}^{\text{nd}} \tilde{M}^{\text{nd}*}
    = -8e^4m^4 \int \frac{\dd p_3^-}{2 \pi} \frac{\dd {p_3'}^-}{2 \pi}
    \exp\left( i\Phi^{\text{dd}} \right)
    \frac{\mcM^{\text{nndd}}}
    {(p_3^2 - m^2 + i \epsilon)({p_3'}^2 - m^2 - i \epsilon)},
\end{equation}
with ${p_3'}^{(+,\perp)} = p_3^{(+,\perp)}$. The phase $\Phi^{\text{dd}}$ reads
\begin{equation}
    \label{eq:phase_dd}
    \Phi^{\text{dd}} = 
    (k_2^- - p_2^-)\delta_2^+
    - (k_1^- - p_1^-)\delta_1^+
    - p_3^-(x_2^+ - x_1^+) 
    + {p_3'}^-(x_2'^+ - x_1'^+) 
    + \Phi_F^{\text{dd}},
\end{equation}
with the field-dependent part $\Phi_F^{\text{dd}}$ given by [we use the
canonical light-cone basis (\ref{eq:canonical_light-cone_basis}) with $P^\mu$
from \eqref{eq:qmu_specified}]
\begin{align}
    &\Phi_F^{\text{dd}} = \phantom{-}\frac{m}{p_3^+}
    \sum\displaylimits_i \xi_i k_1^i \left( 
    \delta_2^+ I_{2i} + \delta_1^+ I_{1i} \right)
    \nonumber\\
    &\mspace{50mu}- \frac{m^2}{p_3^+} \sum\displaylimits_i t_i \xi_i 
    \left( \frac{k_2^+}{p_2^+} \delta_2^+ I_{2i} 
    + \frac{k_1^+}{p_1^+} \delta_1^+ I_{1i} \right)
    - \frac{m^2}{2p_3^+} \sum\displaylimits_i \xi_i^2 
    \left( \frac{k_2^+}{p_2^+} \delta_2^+ J_{2i} 
    + \frac{k_1^+}{p_1^+} \delta_1^+ J_{1i} \right), 
\end{align}
where
\begin{equation}
    I_{ji} = \frac12 \intlim_{-1}^{1} \dd \lambda \,
    \psi_i\left( mX_j^+ + \frac12 m\delta_j^+ \lambda\right),
    \quad
    J_{ji} = \frac12 \intlim_{-1}^{1} \dd \lambda \,
    \psi_i^2\left( mX_j^+ + \frac12 m\delta_j^+ \lambda\right).
\end{equation}
For the products of the noninstantaneous and instantaneous direct terms and vice
versa, we obtain correspondingly
\begin{equation}
    \label{eq:mm_nidd}
    \frac{1}{4} \sum_{\sigma_i, \lambda_i}
    \tilde{M}^{\text{nd}}\tilde{M}^{\text{id}*} 
    = -2 e^4 [2m^2+s(\phi)] \delta(\delta_2^+ + \delta_1^+ - 2 \delta^+)
    \int \frac{\dd p_3^-}{2 \pi}
    \exp\left( i\Phi^{\text{dd}}\right)
    \frac{\mcM^{\text{nidd}}}{p_3^2-m^2+i \epsilon}
\end{equation}
and
\begin{equation}
    \label{eq:mm_indd}
    \frac{1}{4} \sum_{\sigma_i, \lambda_i} 
    \tilde{M}^{\text{id}}\tilde{M}^{\text{nd}*} 
    = -2 e^4 [2m^2+s(\phi)] \delta(\delta_2^+ + \delta_1^+ + 2 \delta^+)
    \int \frac{\dd {p_3'}^-}{2 \pi}
    \exp\left( i\Phi^{\text{dd}}\right)
    \frac{\mcM^{\text{indd}}}{{p_3'}^2-m^2-i \epsilon}.
\end{equation}
Finally, the product of the two instantaneous direct terms is given by
\begin{equation}
    \label{eq:mm_iidd}
    \frac{1}{4} \sum_{\sigma_i, \lambda_i} 
    \tilde{M}^{\text{id}}\tilde{M}^{\text{id}*} 
    = e^4 \delta(\delta_2^+ + \delta_1^+) \delta(\delta^+)
    \exp\left( i\Phi^{\text{dd}}\right) \mcM^{\text{iidd}}.
\end{equation}
The quantities $\mcM^{\text{nndd}}$, $\mcM^{\text{nidd}}$, $\mcM^{\text{indd}}$,
and $\mcM^{\text{iidd}}$ are the traces of the corresponding bispinor parts.
These traces are rearranged with the use of momentum relations in the background
field and subsequently replaced with the rearranged ones in Eqs.
(\ref{eq:mm_nndd}), (\ref{eq:mm_nidd}), (\ref{eq:mm_indd}), and
(\ref{eq:mm_iidd}), which we denote by a tilde: $\mcM^{\text{nndd}} \to
\tilde{\mcM}^{\text{nndd}}$, $\mcM^{\text{nidd}} \to
\tilde{\mcM}^{\text{nidd}}$, etc. Details and explicit expressions are provided
in Appendix~\ref{sec:traces}. The prefactors in Eqs. (\ref{eq:mm_nndd}),
(\ref{eq:mm_nidd}), and (\ref{eq:mm_indd}) are chosen in such a way, that
$\tilde{\mcM}^{\text{nndd}} = \tilde{\mcM}^{\text{indd}} =
\tilde{\mcM}^{\text{nidd}} = 1$ in the limit of a vanishing laser field.

\section{One-step and two-step contributions}
\label{sec:step_split}

As it has been pointed out in the Introduction, in contrast to the vacuum case,
the probability of a tree-level second-order process in an external field [and
hence the cross section (\ref{eq:cross_section_final})] contains contributions
with the intermediate particle being virtual, as well as real, and it can be
written as a sum of so-called one-step and two-step or cascade terms
\cite{baier_higher-order_1972, ritus_vacuum_1972, ilderton_trident_2011,
mackenroth_double_2013, king_trident_2013, dinu_trident_2018, king_trident_2018,
mackenroth_trident_2018, dinu_double_2019}. If the intermediate particle is
real, generally speaking, the propagation distance may be arbitrarily large
inside the field. This causes at least two problems: for sufficiently large
distances, the approximation~(\ref{eq:wave_packets_one-point}) may break down
and also radiative corrections to the electron/photon propagator may become
sizable. On the other hand, in principle, one can recover the two-step
contribution as a combination of the two corresponding first-order processes,
therefore, it is the one-step contribution that is the most nontrivial.

Let us single out the one-step contribution from the cross section
(\ref{eq:cross_section_final}). In our approach, we employ the Schwinger proper
time representation for the denominators of the electron propagators. This
allows us to avoid the use of the Heaviside step functions and to write the
two-step and one-step contributions as integrals with fixed limits. But let us
first highlight the main ideas of the common approach employed in the
literature.

Note that the two-step contribution is contained in the ``nndd'' term
\cite{dinu_trident_2018, dinu_double_2019}. For the ``nndd'' term
(\ref{eq:mm_nndd}), let us consider the integrals in $p_3^-$ and ${p_3'}^-$,
\begin{equation}
    \label{eq:nndd_pminus_init}
    I_{p\indices*{_3},{p_3'}} 
    = \int \frac{\dd p_3^-}{2 \pi} \frac{\dd {p_3'}^-}{2 \pi}
    \frac{\ee^{-ip_3^-(x_2^+ - x_1^+)}\ee^{i{p_3'}^-(x_2'^+ - x_1'^+)}}
    {(p_3^2 - m^2 + i \epsilon)({p_3'}^2 - m^2 - i \epsilon)}.
\end{equation}
Evaluating each of the integrals separately and then combining the results, one
obtains
\begin{equation}
    \label{eq:nndd_pminus_thetas}
    I_{p\indices*{_3},{p_3'}} = \frac{1}{(2p_3^+)^2}
    \exp[-i\tilde{p}_3^-(\delta_2^+ + \delta_1^+)]
    \left[ \theta(p_3^+)\theta(x_2^+-x_1^+)\theta(x_2'^+-x_1'^+)
    + \theta(-p_3^+)\theta(x_1^+-x_2^+)\theta(x_1'^+-x_2'^+) \right].
\end{equation}
The product $\theta(x_2^+-x_1^+)\theta(x_2'^+-x_1'^+)$ can be written as
\cite{dinu_trident_2018, mackenroth_trident_2018}
\begin{equation}
    \label{eq:theta_product}
    \theta(x_2^+-x_1^+)\theta(x_2'^+-x_1'^+) = \theta(\delta^+)\left[1 
    - \theta\left(\frac{|\delta_2^+ + \delta_1^+|}{2} - \delta^+\right)\right].
\end{equation}
In \eqref{eq:theta_product}, a two-step contribution is usually associated with
the first term, and the second term is referred to as a one-step contribution.
Recalling the definition of $\delta^+$ [see \eqref{eq:x_delta_defined}], we
conclude that the function $\theta(\delta^+)$ identifies the two-step
contribution corresponding to the electron emitting a photon first and then
annihilating with the positron into the second photon. Using an analogous
transformation for the product $\theta(x_1^+-x_2^+) \theta(x_1'^+-x_2'^+)$ in
\eqref{eq:nndd_pminus_thetas}, one obtains a two-step contribution $\propto
\theta(-\delta^+)$, which corresponds to the positron emitting a photon first
and then annihilating with the electron into the second photon. The total
two-step contribution can be written as
\begin{equation}
    \label{eq:nndd_pminus_twostep}
    I_{p\indices*{_3},{p_3'}}^{\text{two-step}} = \frac{1}{(2p_3^+)^2}
    \exp[-i\tilde{p}_3^-(\delta_2^+ + \delta_1^+)]
    \theta\left(p_3^+\delta^+\right)
\end{equation}
and the one-step contribution, originating from the ``nndd'' term, as
\begin{align}
    \label{eq:nndd_pminus_onestep}
    I_{p\indices*{_3},{p_3'}}^{\text{one-step}} = -\frac{1}{(2p_3^+)^2} 
    \exp[-i\tilde{p}_3^-(\delta_2^+ + \delta_1^+)]
    \bigg[&\theta\left(p_3^+\right) \theta\left(\delta^+\right) 
    \theta\left(\frac{|\delta_2^+ + \delta_1^+|}{2} - \delta^+\right)
    \nonumber\\&
    + \theta\left(-p_3^+\right) \theta\left(-\delta^+\right)
    \theta\left(\frac{|\delta_2^+ + \delta_1^+|}{2} + \delta^+\right)\bigg].
\end{align}

Now, let us show an alternative way of representing the two-step and one-step
contributions in Eqs. (\ref{eq:nndd_pminus_twostep}) and
(\ref{eq:nndd_pminus_onestep}), respectively. We employ the following
proper-time representation for the denominators:
\begin{equation}
    \frac{1}{p_3^2 - m^2 + i \epsilon} = -i \int\displaylimits_0^\infty \dd s\;
    \ee^{i(p_3^2 - m^2 + i \epsilon)s},
    \quad
    \frac{1}{{p_3'}^2 - m^2 - i \epsilon} = i \int\displaylimits_0^\infty \dd t\;
    \ee^{-i({p_3'}^2 - m^2 - i \epsilon)t}.
\end{equation}
Below, we do not write the terms with $i\epsilon$ for brevity. The integrals in
$p_3^-$ and ${p_3'}^-$ yield [see \eqref{eq:nndd_pminus_init}]
\begin{equation}
    \label{eq:nndd_pminus_done}
    \int \frac{\dd p_3^-}{2 \pi} \frac{\dd {p_3'}^-}{2 \pi}
    \to
    \delta[2p_3^+s - (x_2^+ - x_1^+)] \, 
    \delta[2p_3^+t - (x_2'^+ - x_1'^+)].
\end{equation}
In place of $s$ and $t$, we introduce the variables $\tau$ and $v$
\cite{baier_interaction_1976, meuren_polarization_2013},
\begin{equation}
    \tau = s+t, \quad v = \frac{s-t}{s+t},
    \qquad
    \intlim_0^\infty \dd s \, \dd t 
    \to 
    \intlim_{-1}^1 \dd v \intlim_0^\infty \dd \tau \, \frac{\tau}{2}.
\end{equation}
In terms of the new variables the delta functions in
\eqref{eq:nndd_pminus_done} can be written as
\begin{equation}
    \delta[2p_3^+s - (x_2^+ - x_1^+)] \, 
    \delta[2p_3^+t - (x_2'^+ - x_1'^+)]
    =
    \delta(\delta^+-p_3^+ \tau) \, 
    \delta(\delta_2^+ + \delta_1^+ - 2p_3^+v \tau),
\end{equation}
and the initial quantity $I_{p\indices*{_3},{p_3'}}$ in
\eqref{eq:nndd_pminus_init} reads
\begin{equation}
    \label{eq:nndd_pminus_tau_v}
    I_{p\indices*{_3},{p_3'}} = \intlim_{-1}^1 \dd v \intlim_0^{\infty} \dd\tau\,
    \frac{\tau}{2} \, \delta(\delta^+-p_3^+ \tau) \, 
    \delta(\delta_2^+ + \delta_1^+ - 2p_3^+v \tau)
    \exp[-i\tilde{p}_3^-(\delta_2^+ + \delta_1^+)].
\end{equation}
Evaluating the integrals in $\tau$ and $v$, one obtains that
\begin{equation}
    \label{eq:nndd_pminus_tau_v_done}
    I_{p\indices*{_3},{p_3'}} = \frac{1}{(2p_3^+)^2}
    \exp[-i\tilde{p}_3^-(\delta_2^+ + \delta_1^+)]
    \theta\left(p_3^+\delta^+\right)
    \theta\left( 1 - \left| \frac{\delta_2^+ + \delta_1^+}{2 \delta^+}\right| 
    \right),
\end{equation}
where the first $\theta$-function comes from the integral in $\tau$ and the
second one comes from the integral in $v$. We notice that
\eqref{eq:nndd_pminus_tau_v_done} is the same as \eqref{eq:nndd_pminus_twostep},
apart from the presence of the second $\theta$-function. Then, the two-step
contribution can be written as
\begin{equation}
    \label{eq:nndd_pminus_tau_v_twostep}
    I_{p\indices*{_3},{p_3'}}^{\text{two-step}} 
    = \int \dd v \intlim_0^{\infty} \dd \tau \,
    \frac{\tau}{2} \, \delta(\delta^+-p_3^+ \tau) \, 
    \delta(\delta_2^+ + \delta_1^+ - 2p_3^+v \tau)
    \exp[-i\tilde{p}_3^-(\delta_2^+ + \delta_1^+)],
\end{equation}
which agrees with \eqref{eq:nndd_pminus_twostep} upon the evaluation of the
integrals in $\tau$ and $v$ [note that the limits of the integration in $v$ are
extended to be $(-\infty,\infty)$]. The difference between Eqs.
(\ref{eq:nndd_pminus_tau_v}) and (\ref{eq:nndd_pminus_tau_v_twostep}) is the
one-step contribution,
\begin{equation}
    \label{eq:nndd_pminus_tau_v_onestep}
    I_{p\indices*{_3},{p_3'}}^{\text{one-step}} = - \intlim_{\Gamma_v} \dd v 
    \intlim_0^{\infty} \dd \tau \,
    \frac{\tau}{2} \, \delta(\delta^+-p_3^+ \tau) \, 
    \delta(\delta_2^+ + \delta_1^+ - 2p_3^+v \tau)
    \exp[-i\tilde{p}_3^-(\delta_2^+ + \delta_1^+)],
\end{equation}
with $\Gamma_v = (-\infty,-1) \cup (1, \infty)$. In the
following, we consider the one-step contribution and therefore employ
\eqref{eq:nndd_pminus_tau_v_onestep}. The final expression can be easily
transformed into the result for the two-step contribution
[\eqref{eq:nndd_pminus_tau_v_twostep}] or for the sum of both contributions
[\eqref{eq:nndd_pminus_tau_v}].

\section{Evaluation of the integrals}
\label{sec:integrals}

For the ``nidd'' and ``indd'' terms in Eqs. (\ref{eq:mm_nidd}) and
(\ref{eq:mm_indd}), respectively, we also employ the proper-time representation,
e.g., we have
\begin{equation}
    \label{eq:nidd_pminus_tau}
    \int \frac{\dd p_3^-}{2 \pi} \,
    \frac{\ee^{-ip_3^-(x_2^+ - x_1^+)}}
    {p_3^2-m^2+i \epsilon}
    = -i \int\displaylimits_0^\infty \dd \tau \,
    \frac12 \delta(\delta^+ - p_3^+\tau)
    \exp[-i\tilde{p}_3^-(\delta_2^+ + \delta_1^+)]
\end{equation}
for the ``nidd'' term and an analogous expression for the ``indd'' term [note
that for the ``iidd'' term no proper-time representation is required, since
there are no noninstantaneous parts of the propagators and integrals in the
``--'' momentum components; see \eqref{eq:mm_iidd}]. After that, we notice that
each of the four terms, which we need to compute, contains two delta functions
[see Eqs. (\ref{eq:mm_nndd}), (\ref{eq:mm_nidd}), (\ref{eq:mm_indd}),
(\ref{eq:mm_iidd}), (\ref{eq:nndd_pminus_tau_v_onestep}), and
(\ref{eq:nidd_pminus_tau})], and they allow us to evaluate the integrals in
$\delta^+$ and $\delta_2^+$ in \eqref{eq:cross_section_final}. In place of
$\delta_1^+$ we introduce
\begin{equation}
    \label{eq:rho_defined}
    \rho = \frac{m^2p_2^+}{k_2^+p_3^+}\delta_2^+ 
    + \frac{m^2p_1^+}{k_1^+p_3^+}\delta_1^+
\end{equation}
and we also rescale $\tau$ as
\begin{equation}
    m^2 \tau \to \tau,
\end{equation}
such that the rescaled variable is dimensionless. Then the
direct-direct parts of the total cross section are given by
\begin{align}
    \label{eq:sigma_nndd_prek1}
    \sigma^{\text{nndd}}(\phi) &= \frac{2 \pi^2 r_e^2}{I(\phi)(p_2^++p_1^+)} 
    \intlim_0^{p_2^+ + p_1^+} 
    \frac{\dd k_1^+}{2 \pi} \int \frac{\dd^2 k_1^\perp}{(2 \pi)^2} 
    \intlim_{\Gamma_v} \dd v \intlim_0^\infty \dd \tau \int \dd \rho \,
    \tau \exp\left( i\Phi_v^{\text{dd}}\right) \tilde{\mcM}^{\text{nndd}},
    \\
    \label{eq:sigma_nidd_indd_prek1}
    \sigma^{\text{\{ni\}dd}}(\phi) 
    &= \frac{i \pi^2 r_e^2 [2m^2+s(\phi)]}{2m^2I(\phi)(p_2^++p_1^+)}
    \int\displaylimits_0^{p_2^+ + p_1^+} 
    \frac{\dd k_1^+}{2 \pi} \int \frac{\dd^2 k_1^\perp}{(2 \pi)^2}
    \intlim_0^\infty \dd \tau \int \dd \rho \,
    \left[ \exp\left( i\Phi_1^{\text{dd}}\right) \tilde{\mcM}^{\text{nidd}}
    - \exp\left( i\Phi_{-1}^{\text{dd}}\right) \tilde{\mcM}^{\text{indd}}
    \right],
    \\
    \label{eq:sigma_iidd_prek1}
    \sigma^{\text{iidd}}(\phi) &= -\frac{\pi^2 r_e^2}{I(\phi)(p_2^++p_1^+)} 
    \int\displaylimits_0^{p_2^+ + p_1^+} 
    \frac{\dd k_1^+}{2 \pi} \int \frac{\dd^2 k_1^\perp}{(2 \pi)^2}
    \int \dd \rho \, \exp\left( i\Phi^{\text{iidd}}\right),
\end{align}
where $r_e = \alpha/m$ is the classical electron radius, and the ``nidd'' and
``indd'' terms have been combined as
\begin{equation}
    \sigma^{\text{\{ni\}dd}}(\phi) = \sigma^{\text{indd}}(\phi) +
\sigma^{\text{nidd}}(\phi).
\end{equation}
The phase $\Phi_v^{\text{dd}}$ is given by
\begin{align}
    \label{eq:dd_phase_prek1}
    \Phi_v^{\text{dd}}
    = &-\frac{\rho}{2m^2}(\vec{k}_1^{\perp2} + 2\vec{k}_1^\perp \vec{\mcP}^\perp)
    + \frac{\rho(1+t_1^2+t_2^2)}{4}  \left[ 
    \frac{k_2^{+2}}{p_2^{+2}}(u-1) - \frac{k_1^{+2}}{p_1^{+2}}(u+1) \right]
    \nonumber\\ 
    &+\frac{\rho}{2} \sum_i t_i \xi_i
    \left[ \frac{k_2^{+2}}{p_2^{+2}} (u-1) I_{2i} 
    - \frac{k_1^{+2}}{p_1^{+2}} (u+1) I_{1i} \right]
    +\frac{\rho}{4} \sum_i \xi_i^2
    \left[ \frac{k_2^{+2}}{p_2^{+2}} (u-1) J_{2i} 
    - \frac{k_1^{+2}}{p_1^{+2}} (u+1) J_{1i} \right],
\end{align}
where
\begin{equation}
    \label{eq:nndd_usmall_defined}
    u = \left[\left(\frac{k_2^+}{p_2^+} + \frac{k_1^+}{p_1^+}\right)
    -\frac{4v\tau}{\rho} \right]
    \bigg/\left(\frac{k_2^+}{p_2^+} - \frac{k_1^+}{p_1^+}\right)
\end{equation}
and
\begin{equation}
    \label{eq:dd_phase_pbig}
    \mcP^i = \frac12 m \xi_i \left[ \frac{k_2^+}{p_2^+} (u-1)I_{2i}
    - \frac{k_1^+}{p_1^+} (u+1)I_{1i}\right] - \frac{2mt_iv \tau}{\rho},
\end{equation}
the phases in \eqref{eq:sigma_nidd_indd_prek1} are the same as
$\Phi_v^{\text{dd}}$, but with $v=1$ and $v=-1$, respectively,
and the phase $\Phi^{\text{iidd}}$ is given by
\begin{equation}
    \label{eq:iidd_phase_prek1}
    \Phi^{\text{iidd}} = -\frac{\rho\vec{k}_1^{\perp2}}{2m^2} 
    + \frac{\rho k_2^+k_1^+}{2p_2^+p_1^+}\Big[ 1 + \sum_i (t_i + \xi_i I_i)^2
    + \sum_i \xi_i^2 (J_i - I_i^2) \Big],
\end{equation}
where
\begin{equation}
    I_{i} = \frac12 \intlim_{-1}^{1} \dd \lambda \,
    \psi_i\left( \phi + \frac{k_2^+k_1^+}{2m(p_2^++p_1^+)} \rho \lambda\right),
    \quad
    J_{i} = \frac12 \intlim_{-1}^{1} \dd \lambda \,
    \psi_i^2\left( \phi + \frac{k_2^+k_1^+}{2m(p_2^++p_1^+)} \rho \lambda\right).
\end{equation}
The old variables $\delta_1^+$ and $\delta_2^+$ are expressed via the new ones
as
\begin{equation}
    \label{eq:pegg_nndd_deltas_via_rho}
    \delta_1^+ = \frac{p_3^+ k_1^+}{2m^2p_1^+}(1+u) \rho,
    \quad
    \delta_2^+ = \frac{p_3^+ k_2^+}{2m^2p_2^+}(1-u) \rho.
\end{equation}

The integrals in $\vec{k}_1^\perp$ are Gauss-type (Fresnel) integrals and can be
evaluated analytically [note that the exponential prefactors in Eqs.
(\ref{eq:sigma_nndd_prek1}), (\ref{eq:sigma_nidd_indd_prek1}), and
(\ref{eq:sigma_iidd_prek1}) do not depend on $\vec{k}_1^\perp$; see Appendix
\ref{sec:traces} for details]. However, before being able to perform an integral
in $\vec{k}_1^\perp$, we need to change the order of the integrations and,
strictly speaking, we have to ensure that upon those changes the integrals
remain convergent. It can be seen from Eqs. (\ref{eq:dd_phase_prek1}) and
(\ref{eq:iidd_phase_prek1}) that $\rho=0$ is a possible problematic point. Then,
assuming that, if necessary, the integration contour for $\rho$ is deformed from
$(-\infty,\infty)$ into a new appropriately chosen contour $\Gamma_\rho$ we
obtain that
\begin{equation}
    \label{eq:k1_rho_integral_change}
    \int \frac{\dd^2 k_1^\perp}{(2 \pi)^2} \intlim_{\Gamma_\rho} \dd \rho \, 
    \exp\left[-i\frac{\rho}{2m^2}(\vec{k}_1^{\perp2} 
    + 2\vec{k}_1^\perp \vec{\mcP}^\perp)\right]
    = -\frac{im^2}{2 \pi}\intlim_{\Gamma_\rho} \frac{\dd\rho}{\rho} \, 
    \exp\left(i\frac{\rho \vec{\mcP}^{\perp2}}{2m^2}\right),
\end{equation}
where one should put $\vec{\mcP}^{\perp} = \vec{0}$ for the ``iidd'' term. In
order to specify $\Gamma_\rho$, let us consider the ``iidd'' term and the other
two separately. We start with the ``iidd'' term [\eqref{eq:sigma_iidd_prek1}].

\begin{figure}[t]
    \centering
    \includegraphics{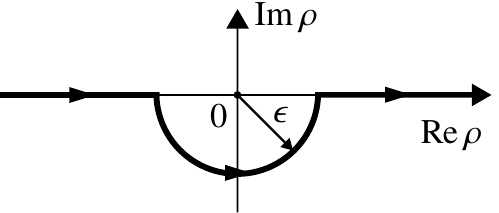}
    \caption{The employed deformation of the integration contour for the 
    ``iidd'' term. As $\epsilon\to0$, the integrals over the two half-lines
    combine together into a principle value integral, and the integral over the
    semicircle results in a term proportional to the residue at $\rho=0$ (see
    the main text for details). The choice of the deformation into the lower
    half-plane is dictated by the fact that after the deformation the integral
    in $\vec{k}_1^\perp$ has to be convergent for a finite value of $\epsilon$.}
    \label{fig:rho_contour}
\end{figure}

If follows from \eqref{eq:k1_rho_integral_change}, that upon the exchange of the
integrations the integral in $\vec{k}_1^\perp$ yields an infinite volume factor,
if $\vec{\mcP}^{\perp} = \vec{0}$ and $\rho = 0$. Therefore, we indeed need to
deform the contour, such that the new contour $\Gamma_\rho$ does not go through
the point $\rho=0$. One of the possibilities is to shift the integration line by
$i \epsilon$ off the real axis. This results in an $i \epsilon$ prescription for
$\rho$ \cite{baier_electromagnetic_1998, dinu_exact_2013}. However, since the
singularity is only at $\rho=0$, it is enough to deform the contour locally by
introducing a semicircle of radius $\epsilon$, as shown in
\figref{fig:rho_contour}. Then, as $\epsilon \to 0$, the integral over the two
half-lines results in the principal value integral, and the integral over the
semicircle yields $i \pi C_{-1}$, with $C_{-1}$ being the residue at $\rho=0$
\cite{ablowitz_complex_2003}.

For the other terms [Eqs. (\ref{eq:sigma_nndd_prek1}) and
(\ref{eq:sigma_nidd_indd_prek1})], the vector $\vec{\mcP}^{\perp}$ is given by
\eqref{eq:dd_phase_pbig}. As a result, upon setting $\rho=0$, the integral in
$\vec{k}_1^\perp$ is evaluated not to an infinite volume factor, but to a delta
function. Therefore, we argue that the deformation of the contour for $\rho$ is
not required for these terms and $\Gamma_\rho = (-\infty,\infty)$. We justify
this by reproducing the vacuum results, known from the literature, if the
external field is set to zero (see below).

We also point out that if one makes the replacement $\rho \to -\rho$, then
\begin{equation}
   \Phi_v^{\text{dd}} \to -\Phi_{-v}^{\text{dd}},
   \quad
   \tilde{\mcM}^{\text{nndd}} \to \tilde{\mcM}^{\text{nndd}}\big|_{v \to -v}.
\end{equation}
Therefore, the integral in $v$ can be reduced to an integral over the interval
$(1,\infty)$ [alternatively, the integral in $\rho$ can be reduced to an
integral over $(0,\infty)$; we use the first option below]. In addition, note
that $\tilde{\mcM}^{\text{indd}} \to \tilde{\mcM}^{\text{nidd}}$ upon the
replacement $\rho \to -\rho$.

As the last steps, we notice that after the integration in $\vec{k}_1^\perp$,
upon rescaling $\rho$ as $\rho k_2^+k_1^+/(p_2^+p_1^+) \to \rho$ for the
``iidd'' term, the integral in $k_1^+$ can be also evaluated analytically and
only a single integral in $\rho$ remains in this term, which can be also written
as an integral over $(0,\infty)$.

\section{Final result}
\label{sec:result}

After all steps described above are carried out, one obtains the final
expressions for the direct-direct contributions to the total cross section [see
\eqref{eq:sigma_dd_terms}],
\begin{align}
    \label{eq:sigma_nndd_final}
    \sigma^{\text{nndd}}(\phi) 
    &= \frac{r_e^2m^2}{I(\phi)(p_2^++p_1^+)}
    \Im \intlim_0^{p_2^++p_1^+} \dd k_1^+ \intlim_1^\infty \dd v 
    \intlim_0^\infty  \dd \tau \int \frac{\dd \rho}{\rho} \, \tau
    \exp\left( i\Phi_v^{\text{dd}}\right) 
    \tilde{\mcM}^{\text{nndd}},
    \\
    \label{eq:sigma_indd_nidd_final}
    \sigma^{\text{\{ni\}dd}}(\phi) 
    &= \frac{r_e^2[2m^2+s(\phi)]}{4I(\phi)(p_2^++p_1^+)}
    \Re \intlim_0^{p_2^++p_1^+} \dd k_1^+
    \intlim_0^\infty \dd \tau \int \frac{\dd \rho}{\rho}
    \exp\left( i\Phi_{1}^{\text{dd}}\right) \tilde{\mcM}^{\text{nidd}},
    \\
    \label{eq:sigma_iidd_final}
    \sigma^{\text{iidd}}(\phi) &= -\frac{r_e^2m^2}{2I(\phi)}
    \left( \intlim_0^\infty \frac{\dd \rho}{\rho} 
    \sin \Phi^{\text{iidd}} + \frac{\pi}{2} \right),
\end{align}
where $\Im$ and $\Re$ denote an imaginary and a real part, respectively,
expressions for the quantities $\tilde{\mcM}^{\text{nndd}}$ and
$\tilde{\mcM}^{\text{nidd}}$ are provided in Appendix \ref{sec:traces}, and the
phase $\Phi_v^{\text{dd}}$ is given by
\begin{align}
    \label{eq:phase_nndd_final}
    \Phi_v^{\text{dd}} = &\frac{\rho}{4}\left[ 
    \frac{k_2^{+2}}{p_2^{+2}} (u-1) - \frac{k_1^{+2}}{p_1^{+2}} (u+1)\right]
    + \frac{\rho}{8} (u^2-1) \sum\displaylimits_i \left(
    \frac{k_2^+}{p_2^+} \zeta_{2i} - \frac{k_1^+}{p_1^+} \zeta_{1i} \right)^2
    \nonumber\\
    &+ \frac{\rho}{4} \sum\displaylimits_i \left[
    \frac{k_2^{+2}}{p_2^{+2}} (u-1) \left(\zeta_{2i}^{(2)} - \zeta_{2i}^2 \right)
    - \frac{k_1^{+2}}{p_1^{+2}} (u+1) 
    \left( \zeta_{1i}^{(2)} - \zeta_{1i}^2 \right) \right],
\end{align}
with
\begin{equation}
    \zeta_{ji} = \frac12 \intlim_{-1}^{1} \dd \lambda \,
    t_i\left(mX_j^+ + \frac12 m \delta_j^+ \lambda\right),
    \quad
    \zeta_{ji}^{(2)} = \frac12 \intlim_{-1}^{1} \dd \lambda \,
    t_i^2\left(mX_j^+ + \frac12 m \delta_j^+ \lambda\right),
\end{equation}
and $X_2^+=x^+ + p_3^+ \tau/(2m^2)$, $X_1^+=x^+ - p_3^+ \tau/(2m^2)$. For the
``iidd'' term, the phase $\Phi^{\text{iidd}}$ is given by
\begin{equation}
    \label{eq:phase_iidd_final}
    \Phi^{\text{iidd}} = \rho\bigg[1 + \sum_i \zeta_i^2
    + \sum_i (\zeta_i^{(2)} - \zeta_i^2)\bigg],
\end{equation}
where
\begin{equation}
    \zeta_i = \frac12 \intlim_{-1}^{1} \dd \lambda \,
    t_i\left(\phi + \frac{p_2^+p_1^+}{m(p_2^++p_1^+)} \rho \lambda\right),
    \quad
    \zeta_i^{(2)} = \frac12 \intlim_{-1}^{1} \dd \lambda \,
    t_i^2\left(\phi + \frac{p_2^+p_1^+}{m(p_2^++p_1^+)} \rho \lambda\right).
\end{equation}

Note that, in order to rewrite the final result via the classical
intensity and the quantum nonlinearity parameters, one needs to simply replace
the ``+'' momentum components with the corresponding quantum nonlinearity
parameters everywhere, except the arguments $\varphi$ of the $t_i(\varphi)$
parameters, where for the general form $\varphi = \phi + \Delta \phi$ of the
argument one also has to multiply $\Delta \phi$ by the factor $m/\xi$ after the
replacement, such that $\Delta \phi \propto 1/\xi$.

\section{Zero-field limit}
\label{sec:zero-field}

In the case of a vanishing plane-wave field, with the use of
Eqs.~(\ref{eq:sigma_nndd_final})--(\ref{eq:sigma_iidd_final}), one should be
able to recover the result known from the literature
\cite{berestetskii_qed_1982}. Since this derivation is different from and also
somewhat less trivial than the one usually presented, we show explicitly how the
vacuum expressions are obtained.

Let us start with the ``iidd'' term in \eqref{eq:sigma_iidd_final}, which is the
simplest out of three. If the external field is set to zero, then
$\Phi^{\text{iidd}} = (1+\vec{t}^{\perp2}) \rho$, where $\vec{t}^{\perp2} =
t_1^2+t_2^2$. The integral in $\rho$ reduces to the Dirichlet integral, and we
obtain that
\begin{equation}
    \label{eq:sigma_iidd_zero}
    \sigma^{\text{iidd}} = - \frac{\pi r_e^2}{4\sqrt{\mu(\mu - 1)}},
\end{equation}
where $\mu$ is the scaled invariant mass squared: $\mu = s/(4m^2)$, with
$s=(p_2+p_1)^2$.

The other two contributions require some more manipulations. Upon setting the
laser field to zero, the quantities $\tilde{\mcM}^{\text{nndd}}$ and
$\tilde{\mcM}^{\text{nidd}}$ are equal to unity, and the phase
$\Phi_v^{\text{dd}}$ reduces to
\begin{equation}
    \label{eq:phase_dd_zero-field}
    \Phi_v^{\text{dd}} = \frac{1}{\rho} + \frac14a^2v^2\tau^2\rho - b v \tau,
\end{equation}
where
\begin{equation}
    \label{eq:ab_defined}
    a = 2 \sqrt{\frac{\vec{t}^{\perp2}(1+\vec{t}^{\perp2})k_2^+k_1^+}
    {p_2^+p_1^+}},
    \quad
    b = \left(\frac{k_2^+}{p_2^+} + \frac{k_1^+}{p_1^+}\right)
    (1 + \vec{t}^{\perp2}),
\end{equation}
and we have rescaled $\rho$ as $\rho/(2\vec{t}^{\perp2}v^2\tau^2) \to \rho$.
After that, the integrals are evaluated in the order shown in Eqs.
(\ref{eq:sigma_nndd_final}) and (\ref{eq:sigma_indd_nidd_final}). Details are
presented in Appendix \ref{sec:zero-field_integrals}. The results are given by
\begin{equation}
    \label{eq:sigma_nndd_zero}
    \sigma^{\text{nndd}} = - \frac{\pi r_e^2}{4 \mu \sqrt{\mu(\mu - 1)}}
\end{equation}
and
\begin{equation}
    \label{eq:sigma_nidd_indd_zero}
    \sigma^{\text{\{ni\}dd}} = \frac{\pi r_e^2}{4 \mu (\mu -1)} 
    \left( \mu + \frac12 \right)
    \ln\left(\frac{\sqrt{\mu}+\sqrt{\mu-1}}{\sqrt{\mu}-\sqrt{\mu-1}}\right),
\end{equation}
where $\ln$ indicates the natural logarithm. Combining all three terms together,
we obtain that
\begin{equation}
    \label{eq:sigma_dd_zero_final}
    \sigma^{\text{dd}} = \frac{\pi r_e^2}{4 \mu^2 (\mu -1)}
    \left[ \mu \left( \mu + \frac12 \right)
    \ln\left(\frac{\sqrt{\mu}+\sqrt{\mu-1}}{\sqrt{\mu}-\sqrt{\mu-1}}\right)
    - (\mu + 1) \sqrt{\mu(\mu - 1)} \right],
\end{equation}
which is the same as the corresponding cross section in Ref.
\cite{berestetskii_qed_1982}. Note that the ``nidd+indd'' term
[\eqref{eq:sigma_nidd_indd_zero}] is the largest and the only positive
contribution to the total cross section (\ref{eq:sigma_dd_zero_final}), and the
``iidd'' term is the largest among the other two by absolute value [compare Eqs.
(\ref{eq:sigma_iidd_zero}) and (\ref{eq:sigma_nndd_zero})].

We point out, that initially the cross section $\sigma^{\text{dd}}$ has been
defined within the light-cone quantization formalism. However, the obtained
expression (\ref{eq:sigma_dd_zero_final}) is the same as the one derived within
the instant-form quantization, which supports the way of defining the cross
section on the light cone, that we have suggested.

Another important remark is the fact that the ``nndd'' term in
\eqref{eq:sigma_nndd_final} does not contain the two-step contribution.
Nevertheless, the complete result has been recovered, which means that the
two-step contribution vanishes in vacuum, as it has to be, if the two-step
contribution indeed corresponds to the physical situation of the intermediate
fermion becoming real. In fact, one can verify this directly by setting the
integration interval for the virtuality $v$ to $(-\infty,\infty)$ and confirming
that the integral vanishes (one should be aware that in this case it is
necessary to recover the $i \epsilon$ prescription for $\tau$ in order to shift
the pole $v=0$ off the real axis).

\section{Discussion of the results}
\label{sec:discussion}

The final result
(\ref{eq:sigma_nndd_final})--(\ref{eq:sigma_iidd_final}) for the total cross
section contains integrals which, generally speaking, have to be evaluated
numerically. Although a numerical analysis of the local cross section is not
given here, let us provide some basic estimates.

We consider the case $\xi \gg 1$ and, for the simplicity of the estimation, we
assume all quantum nonlinearity parameters to be of the order of unity, which is
a regime relevant from the experimental point of view. For this regime, a
general idea is that QED processes in a background field can be described
locally as ones happening in a constant-crossed field (CCF)
\cite{ritus_quantum_effects_1985}. Let us follow this idea and consider the CCF
limit as an approximation to the electron-positron annihilation in the regime of
interest.

In particular, we put $\psi_1(\phi) = \phi$, $\psi_2(\phi)=0$. Then the
phase in \eqref{eq:phase_nndd_final} can be written as
\begin{equation}
    \label{eq:phase_nndd_ccf}
    \Phi_v^{\text{dd}} 
    = \frac{c_{-1}}{\rho} + c_0 + c_1 \rho + c_2 \rho^2 + c_3 \rho^3,
\end{equation}
with the coefficients given by
\begin{equation}
    \label{eq:phase_nndd_ccf_coeffs}
    \begin{aligned}
        c_{-1} &= 2 \vec{t}_F^{\perp2} v^2 \tau^2,
        \\
        c_0 &= -\frac{v \tau (\kappa_2 \chi_1 + \kappa_1 \chi_2)}{\chi_2 \chi_1}
        (1+\vec{t}_F^{\perp2}) 
        - \frac{v^3 \tau^3 (\kappa_2 \chi_1 + \kappa_1 \chi_2)
        (\kappa_2^2 \chi_1^2 + \kappa_1^2 \chi_2^2)}
        {3 \chi_2 \chi_1 (\chi_2 + \chi_1)^2},
        \\
        c_1 &= \frac{\kappa_2 \kappa_1}{2 \chi_2 \chi_1}(1+\vec{t}_F^{\perp2})
        + \frac{v^2 \tau^2 \kappa_2 \kappa_1 (\kappa_2^2 \chi_1^2 
        + \kappa_2 \kappa_1 \chi_2 \chi_1 + \kappa_1^2 \chi_2^2)}
        {2 \chi_2 \chi_1 (\chi_2 + \chi_1)^2},
        \\
        c_2 &= 
        -\frac{v \tau \kappa_2^2 \kappa_1^2 (\kappa_2 \chi_1 + \kappa_1 \chi_2)}
        {4 \chi_2 \chi_1 (\chi_2 + \chi_1)^2},
        \\
        c_3 &= \frac{\kappa_2^3 \kappa_1^3}
        {24 \chi_2 \chi_1 (\chi_2 + \chi_1)^2},
    \end{aligned}
\end{equation}
where
\begin{equation}
    \vec{t}_F^{\perp2} = \left[ t_1(\phi) 
    + \frac{\tau (\kappa_2 \chi_1 + \kappa_1 \chi_2)}{2(\chi_2 + \chi_1)} 
    \right]^2 + t_2^2,
\end{equation}
$t_1(\phi) = t_i + \xi \phi$, $t_i$ are initial values of the parameters
$t_i(\phi)$, $\chi_1$ and $\chi_2$ are the electron and positron quantum
nonlinearity parameters, respectively, and $\kappa_i$ denote the photon
nonlinearity parameters: $\kappa_i = k_i^+ \xi/m$.

The phase in \eqref{eq:phase_iidd_final} is given by
\begin{equation}
     \label{eq:phase_iidd_ccf}
    \Phi^{\text{iidd}} = [1 + \vec{t}^{\perp2}(\phi)] \rho
    + \frac{\chi_2^2 \chi_1^2}{3(\chi_2 + \chi_1)^2} \rho^3,
\end{equation}
where $\vec{t}^{\perp2}(\phi) = t_1^2(\phi) + t_2^2$ in the case of CCF, since
$t_2(\phi) = t_2$.

If $|\vec{t}^{\perp}(\phi)| \lesssim 1$, the influence of an external field on
the annihilation process is expected to be important due to the drastically
different structure of the integrated expressions [compare, e.g., the phases
(\ref{eq:phase_dd_zero-field}) and (\ref{eq:phase_nndd_ccf})]; however, the
total cross section should be of the same order as the one in vacuum, since all
parameters are less or of the order of unity. Note that in this regime one might
need to take into account the term corresponding to the interference of the
direct and the exchange diagrams, which has not been discussed here.

Moreover, for the regime $|\vec{t}^\perp(\phi)| \lesssim 1$ one should also keep
in mind that electron-positron annihilation into one photon becomes dominant for
sufficiently small values of $|\vec{t}^\perp(\phi)|$. In fact, the annihilation
into one photon is a resonant process, the local cross section reaches its
highest value (which might be of the order of $r_e^2/\alpha$) as
$|\vec{t}^\perp(\phi)| \to 0$ and becomes exponentially suppressed as
$|\vec{t}^\perp(\phi)|$ grows \cite{ritus_quantum_effects_1985}.

On the other hand, typical values of $|\vec{t}^\perp(\phi)|$ are much
larger than unity for sufficiently long laser pulses [see the discussion below
\eqref{eq:inv_mass_sqr_field}]. Therefore, let us consider the regime
$|\vec{t}^{\perp}(\phi)| \gg 1$, where the obtained cross section is expected to
be dominant.

Let us estimate the formation regions of the integrals in $\rho$ and
$\tau$ for Eqs.~(\ref{eq:sigma_nndd_final}), (\ref{eq:sigma_indd_nidd_final})
and in $\rho$ for  \eqref{eq:sigma_iidd_final}. We start with
Eqs.~(\ref{eq:sigma_nndd_final}) and~(\ref{eq:sigma_indd_nidd_final}), i.e.,
with the phase (\ref{eq:phase_nndd_ccf}). We assume that at least for the major
part of the parameter space under consideration the equations $\partial
\Phi_v^{\text{dd}}/\partial \rho = 0$ and $\partial \Phi_v^{\text{dd}}/\partial
\tau = 0$ are not satisfied simultaneously for $\rho \in \mathbb{R}$, $\tau >
0$, and $v \geq 1$. Then the integral in $\rho$ is formed around stationary
points (we expect the equation $\partial \Phi_v^{\text{dd}}/\partial \rho = 0$
to have at least two real roots), and the integral in $\tau$ is formed around
zero. From \eqref{eq:phase_nndd_ccf_coeffs} it follows that $|c_0| \sim 1$ when
$v \tau \vec{t}^{\perp2}(\phi) \sim 1$. Then we estimate that the integral in
$\tau$ forms at the interval corresponding to $v \tau \lesssim
1/\vec{t}^{\perp2}(\phi)$. Comparing terms with different powers of $\rho$ in
$\partial \Phi_v^{\text{dd}}/\partial \rho$, one concludes that for $\rho$ the
formation region can be estimated as $|\rho| \lesssim 1/\vec{t}^{\perp2}(\phi)$.
Note that due to a possible cancellation of terms in the phase, the actual
scaling might change, e.g., for specific values of $\kappa_1$, however, we
assume that the formation regions still decrease with the growth of
$|\vec{t}^{\perp}(\phi)|$ fast enough, such that the following considerations
for the total cross section are valid.

From \eqref{eq:phase_iidd_ccf} one concludes that the formation region for
$\rho$ is the same as for \eqref{eq:phase_nndd_ccf}.

If $v \tau, |\rho| \lesssim 1/\vec{t}^{\perp2}(\phi) \ll 1$, one can
neglect terms $\ll 1$ in \eqref{eq:phase_nndd_ccf} in such a way that
\eqref{eq:phase_nndd_ccf} is reduced to the phase in vacuum, i.e., to
\eqref{eq:phase_dd_zero-field} (after rescaling
$\rho/[2\vec{t}^{\perp2}(\phi)v^2 \tau^2] \to \rho$), with the change
$\vec{t}^\perp \to \vec{t}^\perp(\phi)$. Equation (\ref{eq:phase_iidd_ccf}) can
be transformed analogously. Then one finds that \eqref{eq:sigma_indd_nidd_final}
produces the leading contribution and the total cross section can be estimated
as $\sigma^{\text{dd}}(\phi) \sim r_e^2 \mu^{-1}(\phi) \ln \mu(\phi)$, where
$\mu(\phi) = s(\phi)/4m^2$ and $s(\phi)$ is given by
\eqref{eq:inv_mass_sqr_field}. This result is the same as for the analogous
regime $\mu \gg 1$ in vacuum, with the change $\mu \to \mu(\phi)$ [see
\eqref{eq:sigma_dd_zero_final}].

Transferring the estimates, obtained for CCF, to the annihilation process in a
general field of extreme intensity, we conclude that, under the aforementioned
assumptions, high-intensity background fields are not expected to increase or
suppress the cross section by orders of magnitude in comparison to the one in
vacuum, if the local parameters of the collision are similar in both cases.
However, if the external field significantly alters the initial value
$\vec{t}^\perp$ of the parameter $\vec{t}^\perp(\phi)$, the overall result might
differ considerably from the vacuum one for particular field configurations and
collision geometries.

For a quantitative estimate of the importance of the reaction $e^-e^+
\Rightarrow \gamma\gamma$, let us assume a quasineutral electron-positron plasma
with the density $\varrho$. The relative change of the density due to the
annihilation can be estimated as $\Delta \varrho/\varrho \sim \sigma \varrho l$,
where $\sigma$ and $l$ are the typical cross section and the length scale of
interaction, respectively. Let us take $\sigma \sim
r_e^2/\langle\mu(\phi)\rangle$, with $\langle\mu(\phi)\rangle \sim \xi^2 \sim
10^2$ (where we put $\xi \sim 10$), solid-state density $\varrho \sim 10^{23}$
$\text{cm}^{-3}$, and $l \sim 100$~$\mu\text{m}$. We obtain that $\Delta
\varrho/\varrho \sim 10^{-6}$. If instead we take $\sigma \sim r_e^2$ (an
estimate for the regime $|\vec{t}^{\perp}(\phi)| \lesssim 1$), the result is
$\Delta \varrho/\varrho \sim 10^{-4}$, which is still a small relative change.
This implies that electron-positron annihilation into two photons is not
significant for current-technology laser-based experiments.

\section{Conclusions}
\label{sec:conclusions}

We have investigated analytically the process of annihilation of an
electron-positron pair into two photons in the presence of an intense plane-wave
field, as a characteristic example of $2\Rightarrow2$ reactions. The external
field has been taken into account exactly in the calculations by working in the
Furry picture, and light-cone quantization has been employed, in order to have a
formalism particularly suitable for studying processes in a plane-wave
background field.

Though the presented description of the scattering based on the use of wave
packets is tailored to the reaction $e^-e^+ \Rightarrow \gamma\gamma$ in a laser
pulse, it applies to a general second-order 2-to-2 reaction in an intense
background field. We have seen that it is convenient to introduce the concept of
a local cross section, which although not being a measurable quantity, is a
useful tool especially for comparison of the results in a laser field and the
corresponding ones in vacuum. Indeed, the local cross section in a plane-wave
field is a qualitatively different entity with respect to its vacuum limit,
since it bears the dependence on the light-cone moment of the collision and may
also become negative in some regions of the parameter space. Therefore, the
cross section in the external field cannot be seen as an observable, but instead
could be interpreted as a quantity, which extends the concept of the classical
cross section, similar to the relation between the Wigner distribution and the
classical phase-space distribution.

In contrast to processes in a plane wave initiated by a single particle, the
pair annihilation into two photons does also occur in vacuum. The vacuum part
has an additional momentum-conserving delta function at each vertex, which is
hidden, if one works in the Furry picture (see Ref.
\cite{ilderton_analytic_2020} for a discussion of splitting the amplitude of a
second-order tree-level process in a laser field into different parts). Our
definition of the cross section and also the analytical evaluation of Gauss-type
integrals in the transverse momentum components of the final particles
effectively remove those delta functions and allow one to write the total local
cross section without a formal split into a vacuum and a field-dependent part.
We have also ensured that by setting the external field to zero, the vacuum
cross section is recovered.

A distinct feature of second-order tree-level processes in an intense background
is a nonvanishing contribution from the cascade or two-step channels, which
correspond to the intermediate particle becoming real. In contrast to 1-to-3
reactions, 2-to-2 reactions have not one, but two cascade channels, which in the
case of $e^-e^+ \Rightarrow \gamma\gamma$ correspond to either the electron or
the positron emitting first a photon and then annihilating with the other
particle into the second photon. Though the different contributions can be
treated in a standard fashion, which involves the use of Heaviside step
functions, we have demonstrated a concise way of representing them via
virtuality integrals with fixed integration limits.

We have explicitly evaluated the total cross section, without taking into
account the interference term between the direct and the exchange amplitudes. In
addition to the common classical nonlinearity  and quantum nonlinearity
parameters $\xi$ and $\chi_i$, respectively, the final result depends
nontrivially on the parameters $t_i$ (similar to the annihilation into one
photon \cite{ritus_quantum_effects_1985, tang_one-photon_2019}), which can be
related to transverse momentum components of the incoming particles with respect
to the laser pulse propagation direction. One can distinguish two regimes
$|\vec{t}^\perp(\phi)| \lesssim 1$ and $|\vec{t}^\perp(\phi)| \gg 1$ depending
on the magnitude of the local quantity $|\vec{t}^\perp(\phi)| =
\sqrt{t_1^2(\phi) + t_2^2(\phi)}$, where $\phi$ is the laser phase at the
collision point $x^\mu$.

In the highly nonlinear case ($\xi \gg 1$), if dynamics over several laser
periods is considered, typical values of $|\vec{t}^\perp(\phi)|$ are much larger
than unity. For $|\vec{t}^\perp(\phi)| \gg 1$ the cross section
$\sigma^{\text{dd}}(\phi)$ presented here should account for the most
significant contribution to electron-positron annihilation. The considerations
for the constant-crossed field limit imply that for $\xi \gg 1$, $\chi_i \sim
1$, $|\vec{t}^\perp(\phi)| \gg 1$ the cross section in a background field
behaves analogously to the cross section in vacuum (to leading order), with the
replacement of the asymptotic invariant mass $\sqrt{s}$ with its local value
$\sqrt{s(\phi)}$. This suggests that the cross section in an intense field is
similar in magnitude to the one in vacuum, if the local parameters of the
collision are the same. However, due to the change of $|\vec{t}^\perp(\phi)|$
with the laser phase $\phi$, the average effect of the presence of the field
might be considerable.

Finally, simple numerical estimates indicate that electron-positron annihilation
into two photons is not sizable in current laser-based experiments. However, it
might play an important role in other setups, e.g., in an astrophysical
environment, where the length scales of interaction are very large.

\section*{Acknowledgments}
The authors acknowledge useful discussions with Victor Dinu, Gregor Fauth, and
Christoph Keitel. SB would also like to thank Oleg Skoromnik, Alessandro Angioi,
Stefano Cavaletto, Ujjwal Sinha, Halil Cakir, Petr Krachkov, Salvatore
Castrignano, Archana Sampath, Maitreyi Sangal, Matteo Tamburini, Matthias
Bartelmann, Daniel Bakucz Can\'{a}rio, Sergei Kobzak, Dominik Lentrodt, Calin
Hojbota, and Dmitry Zhukov for valuable discussions and suggestions.

\appendix

\section{Light-cone quantization}
\label{sec:light-cone_quantization}

We define the light-cone coordinates in a covariant way using the light-cone
basis $\{\eta^\mu, \bar{\eta}^\mu, e_1^\mu, e_2^\mu\}$, with the four-vectors of
this basis satisfying the following properties \cite{meuren_polarization_2013}:
\begin{equation}
    \label{eq:coordinates_general_properties}
    \eta^2 = \bar{\eta}^2 = 0,
    \quad
    \eta \bar{\eta} =1,
    \quad
    \eta e_i = \bar{\eta} e_i = 0,
    \quad
    e_i e_j = - \delta_{ij}.
\end{equation}
Then an arbitrary four-vector $a^\mu$ can be written as
\begin{equation}
    a^\mu = a^+ \bar{\eta}^\mu + a^- \eta^\mu + a^1 e_1^\mu + a^2 e_2^\mu,
\end{equation}
where
\begin{equation}
    \label{eq:light-cone_components_defined}
    a^+ = a \eta,
    \quad
    a^- = a \bar{\eta},
    \quad
    a^1 = -ae_1,
    \quad
    a^2 = -ae_2.    
\end{equation}
The metric tensor is given by
\begin{equation}
    g^{\mu \nu} = \eta^\mu \bar{\eta}^\nu + \bar{\eta}^\mu \eta^\nu 
    - e_1^\mu e_1^\nu - e_2^\mu e_2^\nu,
\end{equation}
which can be written in the matrix form as
\begin{equation}
    g^{\mu \nu} =
    \begin{pmatrix}
        0 & 1 & \phantom{-}0 & \phantom{-}0 \\
        1 & 0 & \phantom{-}0 & \phantom{-}0 \\
        0 & 0 & -1 & \phantom{-}0 \\
        0 & 0 & \phantom{-}0 & -1
    \end{pmatrix}
\end{equation}
(note that the order of the components is +, --, 1, 2).  The scalar product of
two four-vectors $a^\mu$ and $b^\mu$ is
\begin{equation}
    ab = a^+b^- + a^-b^+ + a^ib_i = a^+b^- + a^-b^+ 
    - \boldsymbol{a}^{\perp} \boldsymbol{b}^{\perp},
\end{equation}
where $\boldsymbol{a}^{\perp} = (a^1, a^2)$ and $\boldsymbol{b}^{\perp} = (b^1,
b^2)$. For the quantization in the presence of a plane-wave field $A^\mu(x) =
A^\mu(k_0x)$, we choose $\eta^\mu = k_0^\mu/m$. We also need to fix the signs of
scalar products. In order to do that, we assume the signature $(+,-,-,-)$ for
the metric tensor in the instant form. Then, we have $p^+>0$, $p^2=m^2$ for an
on-shell fermion with four-momentum $p^\mu$.

The derivation of the light-front Hamiltonian is analogous to the one in the
vacuum case (see Refs.~\cite{mustaki_perturbative_1991, brodsky_quantum_1998,
burkardt_light-front_2002}); however, the background field $A^\mu(k_0x)$ is
included in the zeroth-order Hamiltonian $H_0$ \cite{furry_bound_1951,
fradkin_unstable_1991}. The result is \cite{mustaki_perturbative_1991,
brodsky_quantum_1998, burkardt_light-front_2002, bragin_dissertation_2019}
\begin{equation}
    \label{eq:hamiltonian}
    H = H_0 + V_1 + V_2 + V_3
\end{equation}
with
\begin{equation}
    \label{eq:hamiltionian_terms}
    \begin{aligned}
        &H_0 = \int \dd^2 x^\perp \dd x^-\, 
        \Big[ \overline{\psi}\gamma^- i\partial_- \psi 
        + e\overline{\psi}\gamma^+ \psi A^-  
        + \frac12 (\partial_- \mcA^-)^2 
        + \frac12 (\partial_1 \mcA_2 - \partial_2 \mcA_1)^2 \Big],
        \\
        &V_1 = e \int \dd^2 x^\perp \dd x^-\, 
        \overline{\psi}\gamma^\mu \psi \mcA_\mu,
        \\
        &V_2 = \frac{e^2}{2} \int \dd^2 x^\perp \dd x^-\,
        \mcA_\mu \overline{\psi} \gamma^\mu \frac{\gamma^+}{i\partial_-}
        \gamma^\nu \psi \mcA_\nu,
        \\
        &V_3 = \frac{e^2}{2} \int \dd^2 x^\perp \dd x^-\,
        \overline{\psi} \gamma^+ \psi \frac{1}{(i\partial_-)^2} 
        \overline{\psi} \gamma^+ \psi,
    \end{aligned}
\end{equation}
where $\psi$ and $\mcA^\mu$ are the electron and photon fields, respectively, to
be quantized (in fact, only the projection $\psi_+ = \Lambda_+\psi$ is an
independent degree of freedom, where $\Lambda_+ = \gamma^- \gamma^+/2$, and
$\mcA^\mu$ has only two independent components \cite{mustaki_perturbative_1991,
brodsky_quantum_1998}).

The Dirac equation for the electron field $\psi$ is
$[\gamma(i\partial-eA)-m]\psi=0$, as a result, in the interaction picture we
obtain the following expansion of $\psi(x)$ via the Volkov wave functions (see
Refs.~\cite{bergou_wavefunctions_1980,bragin_dissertation_2019} for discussions
of the completeness of the Volkov solutions on the light cone):
\begin{equation}
    \psi(x) = \sum_{\sigma} 
    \int \frac{\tilde{\dd}^3p}{(2\pi)^3}
    \left[ a\indices*{_{p \sigma}} \psi_{p \sigma}(x) 
    + b_{p \sigma}^\dagger \psi_{p \sigma}^{(-)}(x) \right],
\end{equation}
where
\begin{equation}
    \frac{\tilde{\dd}^3 p}{(2 \pi)^3} = \frac{\dd^2 p^\perp}{(2 \pi)^2}
    \frac{\dd p^+}{2 \pi} \theta(p^+),
\end{equation}
$a\indices*{_{p \sigma}}$, $b\indices*{_{p \sigma}}$ ($a_{p \sigma}^\dagger$,
$b_{p \sigma}^\dagger$) are the annihilation (creation) operators, with the
anticommutation relations
\begin{equation}
    \big\{ a\indices*{_{p \sigma}}, a_{p' \sigma'}^\dagger \big\} 
    = \big\{ b\indices{_{p \sigma}}, b_{p' \sigma'}^\dagger \big\} 
    = (2\pi)^3 \delta^{(+,\perp)}(p - p') \delta_{\sigma \sigma'},
\end{equation}
$\psi_{p \sigma}(x)$ are the positive-energy Volkov wave functions
(\ref{eq:volkov_solution_positive}), and $\psi_{p \sigma}^{(-)}(x)$ are the
negative-energy ones,
\begin{equation}
    \label{eq:volkov_solution_negative}
    \psi_{p \sigma}^{(-)}(x) 
    = \frac{K_{-p}(\phi)v_{p \sigma}}{\sqrt{2p^+}} \ee^{iS_{-p}(x)}, 
\end{equation}
with the free Dirac bispinor $v_{p \sigma}$ defined such that
$\bar{v}_{p\sigma}v_{p \sigma'} = -2m\delta_{\sigma\sigma'}$,
$\bar{v}_{p\sigma}\gamma^\mu v_{p\sigma'} = 2p^\mu\delta_{\sigma\sigma'}$,
$\sum_\sigma v_{p\sigma}\bar{v}_{p\sigma} 
= \gamma p - m$~\cite{berestetskii_qed_1982}.

The quantized part $\mcA^{\mu}(x)$ of the photon field  is represented in the
same way, as in the vacuum case \cite{bjorken_quantum_1971,
mustaki_perturbative_1991, brodsky_quantum_1998},
\begin{equation}
    \mcA^\mu(x) = \sum_{\lambda} 
    \int \frac{\tilde{\dd}^3k}{(2\pi)^3}
    \left[ c\indices*{_{k \lambda}} \phi_{k \lambda}^{\mu}(x)  
    + c_{k \lambda}^\dagger \phi_{k \lambda}^{*\mu}(x) \right],
\end{equation}
where the creation and annihilation operators obey the relation
\begin{equation}
    \big[ c\indices{_{k \lambda}}, c_{k' \lambda'}^\dagger\big] 
    = (2\pi)^3 \delta^{(+,\perp)}(k - k') 
    \delta_{\lambda \lambda'},
\end{equation}
and $\phi_{k \lambda}^{\mu}(x)$ is given by
\begin{equation}
    \phi_{k \lambda}^{\mu}(x) = \frac{\epsilon^\mu_{k \lambda}}{\sqrt{2k^+}}
    \ee^{-ikx},
\end{equation}
with the polarization four-vectors $\epsilon^\mu_{k \lambda}$ satisfying the
conditions
\begin{equation}
    \epsilon^\mu_{k \lambda} \epsilon^*_{k \lambda' \mu} 
    = - \delta_{\lambda \lambda'},
    \quad
    k\indices*{_\mu} \epsilon^\mu_{k \lambda} = 0,
    \quad
    \sum_{\lambda} \epsilon^\mu_{k \lambda} \epsilon^{*\nu}_{k \lambda}
    = -g^{\mu \nu} + \frac{\eta^\mu k^\nu + \eta^\nu k^\mu}{k^+}.
\end{equation}

\section{Wave packets}
\label{sec:wave_packets}

A positive-energy wave packet $\Psi_p(x)$ with the central four-momentum $p^\mu$
(the polarization degree of freedom is suppressed) is constructed according to
\eqref{eq:wave_packet_definition}. The density $\tilde{f}_p(q)$ is defined such
that $\Psi_p(x)$ is normalized to one particle,
\begin{equation}
    \label{eq:wave_packet_normalization}
    \int \dd^2 x^\perp \dd x^- J_e^+(x)
    = \int \frac{\tilde{\dd}^3 q}{(2 \pi)^3} |\tilde{f}_p(q)|^2 = 1.
\end{equation}
The four-current density is defined as $J_e^\mu(x) = \overline{\Psi}_p(x)
\gamma^\mu \Psi_p(x)$ \cite{berestetskii_qed_1982}. By assuming that
$\tilde{f}_p(q)$ is peaked around the four-momentum $p^\mu$ and by taking into
account that the bispinor part of the wave packet is slowly varying with
$q^\mu$, we obtain that
\begin{equation}
    \label{eq:current_plane_wave}
    J_e^\mu(x) \approx |f_p(x)|^2 \frac{\pi_{e,p}^\mu(\phi)}{p^+},
\end{equation}
where $f_p(x)$ is given by \eqref{eq:wave_packet_position}, the subscript $e$
denotes the electron current density, and $\pi_{e,p}^\mu(\phi) =
\pi_p^\mu(\phi)$ [see \eqref{eq:dressed_momentum}].

For a positron with the wave-packet density $\tilde{f'}_{p}^*(q)$ (an asterisk
denotes the complex conjugate), one obtains that
\begin{equation}
    J_{-e}^{\mu}(x) \approx |f'_p(x)|^2 \frac{\pi_{-e,p}^{\mu}(\phi)}{p^+},
\end{equation}
where
\begin{equation}
    \label{eq:dressed_momentum_positron}
    \pi_{-e,p}^{\mu}(\phi) = -\pi_{-p}^{\mu}(\phi).
\end{equation}

Physically, the quantity $|f_p(x)|^2$ (and $|f'_p(x)|^2$ for positrons) has a
particularly transparent form in the considered case of a narrow wave packet in
momentum space. We indicate as (we focus on electrons, for positrons all
considerations are analogous)
\begin{equation}
    h_p(x)=\int \frac{\tilde{\dd}^3 q}{(2 \pi)^3} \tilde{f}_p(q) \exp(-iqx)
\end{equation}
the asymptotic form of \eqref{eq:wave_packet_position} for $\phi\to-\infty$,
where the field-dependent part of the phase vanishes. By expanding the phase
$qx$ up to leading order in $\vec{q}^{\perp}-\vec{p}^{\perp}$ and $q^+-p^+$, one
neglects the spreading of the wave packet, and it is easy to see that if the
function $h_p(x)$ is peaked at $x^+=0$ around the point
$\vec{x}^{\perp}=\vec{0}$ and $x^-=0$, then for a generic $x^+$ it will be
peaked at $\vec{x}^{\perp}=\vec{p}^{\perp}x^+/p^+$ and
$x^-=(m^2+\vec{p}^{\perp2})x^+/2p^{+2}$, i.e., it will follow the free classical
trajectory. By carrying out the same calculation with the full wave packet
$f_p(x)$ [see \eqref{eq:wave_packet_position}], one obtains
\begin{equation}
    \label{F_approx}
    |f_p(x)|^2=|F_p(x)|^2 \approx 
    \left|\int \frac{\tilde{\dd}^3 q}{(2 \pi)^3} \tilde{f}_p(q) \exp[-i(q-p)x 
    - i\vec{\nabla}_{\vec{p}^{\perp}}\mcS_p(\phi)(\vec{q}^{\perp}-\vec{p}^{\perp}) 
    - i\partial_{p^+}\mcS_p(\phi)(q^+-p^+)]\right|^2.
\end{equation}
Now, by recalling that the phase of a positive-energy Volkov state corresponds
to the classical action of an electron in the corresponding plane wave, one
obtains that $|f_p(x)|^2 \approx|h_p(x_p)|^2$, where
$x_p^{\mu}=[0,x^-_p(x^+),\vec{x}^{\perp}_p(x^+)]$, with
\begin{align}
    x^-_p(x^+)&=x^--\frac{m^2+\vec{p}^{\perp2}}{2p^{+2}}x^+
    +\intlim_{-\infty}^{\phi}\dd\beta
    \left(\frac{e\vec{p}^{\perp}\vec{A}^{\perp}(\beta)}{mp^{+2}}
    -\frac{e^2\vec{A}^{\perp2}(\beta)}{2mp^{+2}}\right),
    \\
    \vec{x}^{\perp}_p(x^+)&=\vec{x}^{\perp}-\frac{\vec{p}^{\perp}}{p^+}x^+
    +\frac{e}{mp^+}\intlim_{-\infty}^{\phi}\dd\beta\vec{A}^{\perp}(\beta),
\end{align}
which indicates that the function $|f_p(x)|^2$ is centered around the classical
trajectory of the electron in the plane wave under consideration.

\section{Conditions for the approximations for the wave packets}
\label{sec:conditions}

Here we provide a discussion about the approximations given in Eqs.
(\ref{eq:wave_packet_approximation1}) and (\ref{eq:wave_packets_one-point}).

For the case of a plane-wave field $A^\mu(k_0x)$, since the dependence of the
field on $x^\mu$ is only via the light-cone time $x^+$, the conditions for the
approximations for the components $x^-$, $\vec{x}^\perp$, that one needs to make
in order to obtain the final expression (\ref{eq:w_total_one-point}), are
ultimately the same as in vacuum, i.e., related only to the resolution of the
detector and the widths of the wave packets.

In order to understand the conditions for the light-cone time in
\eqref{eq:wave_packet_approximation1}, i.e., for the variable $\delta_1^+$, let
us consider the approximation $F_1(X_1-\delta_1/2) \approx F_1(X_1)$ (for the
positron wave packet the considerations below proceed analogously). Let us
assume to work in the highly nonlinear regime, i.e., $\xi \gg 1$. Requiring the
correction to the phase in \eqref{eq:wave_packet_amplitude} due to $\delta_1^+$
to be small, and keeping only linear terms in $\delta_1^+$ and in the widths
$\Delta p_1^+$ and $\Delta \vec{p}_1^\perp$ of the wave packet, one arrives at
the following condition:
\begin{equation}
    \left| 
    \frac{\vec{\pi}_{p_1}^{\perp}(\Phi_1)\Delta\vec{p}_1^{\perp}}{2p_1^+}\delta_1^+
    - \frac{m^2+\vec{\pi}_{p_1}^{\perp2}(\Phi_1)}{4p_1^{+2}}\Delta p_1^+\delta_1^+
    \right| \ll 1,
\end{equation}
where $\Phi_1 = mX_1^+$. Considering each term separately, we obtain
that the conditions on the widths of the wave packet are
\begin{equation}
    \label{eq:conditions_formal}
    |\Delta \vec{p}_1^\perp| \ll 
    \frac{p_1^+}{|\vec{\pi}_{p_1}^{\perp}(\Phi_1)|| \delta_1^+|},
    \quad
    \Delta p_1^+ \ll 
    \frac{p_1^{+2}}{[m^2+\vec{\pi}_{p_1}^{\perp\,2}(\Phi_1)]| \delta_1^+|}.
\end{equation}

The approximation in \eqref{eq:wave_packets_one-point} is qualitatively
different than that in \eqref{eq:wave_packet_approximation1}, since it is an
approximation for the particle densities (which are classical concepts), rather
than for the wave packets themselves. However, it can be related to the
approximation (\ref{eq:wave_packet_approximation1}), since the
conditions for the approximation $|f_1(x- \delta/2)|^2 \approx |f_1(x)|^2$ can
be written as in \eqref{eq:conditions_formal}, with the replacements $\Phi_1 \to
\phi$, $\delta_1^+ \to \delta^+$ [due to $|f_1(x-\delta/2)|^2 =
|F_1(x-\delta/2)|^2$]. Then, if $|\delta^+| \lesssim |\delta_1^+|$ and
$|\vec{\pi}_{p_1}^{\perp}(\phi)| \lesssim |\vec{\pi}_{p_1}^{\perp}(\Phi_1)|$ the
approximations (\ref{eq:wave_packet_approximation1}) and
(\ref{eq:wave_packets_one-point}) are valid simultaneously, when
\eqref{eq:conditions_formal} is fulfilled.

In order to assess the magnitude of $\delta_1^+$, $\delta_2^+$, and
$\delta^+$, let us employ the ideas, presented in Sec.~\ref{sec:discussion}. For
simplicity, let us also use the canonical light-cone basis
(\ref{eq:canonical_light-cone_basis}) with \eqref{eq:qmu_specified}, then
$\vec{\pi}_{p_1}^\perp(\phi) = m \vec{t}^\perp(\phi)$,
$\vec{\pi}_{p_1}^\perp(\Phi_1) = m \vec{t}^\perp(\Phi_1)$. We consider the
one-step contribution, therefore $v \geq 1$.

If $|\vec{t}^\perp(\phi)| \lesssim 1$, the integrals are expected to
form at $|\rho|$, $v\tau \lesssim 1$ (note that here and below, like in
Sec.~\ref{sec:discussion}, we assume the quantum nonlinearity parameters to be
of the order of unity). Therefore, $|\delta_1^+|$, $|\delta_2^+|$, $|\delta^+|
\lesssim 1/(m\xi)$ [see \eqref{eq:pegg_nndd_deltas_via_rho} and also note that
$\delta^+ = p_3^+ \tau/m^2$], analogously to first-order processes
\cite{ritus_quantum_effects_1985}. For $|\vec{t}^{\perp}(\Phi_1)|$, let us take
the upper bound $|\vec{t}^{\perp}(\Phi_1)| \sim 1$. Then from
\eqref{eq:conditions_formal} we obtain
\begin{equation}
    |\Delta \vec{p}_1^\perp| \ll m,
    \quad
    \Delta p_1^+ \ll p_1^+.
\end{equation}

If $|\vec{t}^\perp(\phi)| \gg 1$, the formation regions are defined by
$|\rho|$, $v\tau \lesssim 1/\vec{t}^{\perp2}(\phi)$, then $|\delta_1^+|$,
$|\delta_2^+|$, $|\delta^+| \lesssim 1/[m\xi\vec{t}^{\perp2}(\phi)]$. From
\eqref{eq:conditions_formal} it follows that
\begin{equation}
    |\Delta \vec{p}_1^\perp| \ll m |\vec{t}^{\perp}(\phi)|,
    \quad
    \Delta p_1^+ \ll p_1^+,
\end{equation}
where we have employed the fact that $|\vec{t}^{\perp}(\Phi_1)| \approx
|\vec{t}^{\perp}(\phi)|$ for $\tau \lesssim 1/\vec{t}^{\perp2}(\phi)$ (recall
that $v \geq 1$).

Combining both cases together, we write the relations
(\ref{eq:conditions_formal}) as
\begin{equation}
    \label{eq:conditions_final}
    |\Delta \vec{p}_1^\perp| \ll \max(m, |\vec{\pi}_{p_1}^{\perp}(\phi)|),
    \quad
    \Delta p_1^+ \ll p_1^+.
\end{equation}

Under the conditions (\ref{eq:conditions_final}) (and analogous ones for
the positron wave packet), both approximations
(\ref{eq:wave_packet_approximation1}) and (\ref{eq:wave_packets_one-point}) can
be employed for the one-step contribution.

As to the two-step contribution, the integration limits for $v$ are
$(0,1)$ instead of $(1,\infty)$ in \eqref{eq:sigma_nndd_final}. Then the product
$v \tau$ can be made arbitrarily small even for $\tau \gg 1$. This hints that
$m\delta^+$ might in principle be of the order of the total laser phase. Hence,
if we are to employ the cross section (\ref{eq:cross_section_final}), in
general, we need to restrict ourselves to the evaluation of the one-step
contribution alone, unless we consider incoming wave packets, which are broader
in configuration space than the laser pulse.

We emphasize the semiquantitative nature of the above considerations and point
out the importance of performing the (numerical) evaluation with the wave
packets in order to ascertain precisely the conditions under which the approach
based on the local cross section in Eq. (\ref{eq:cross_section_final}) is
applicable.

\section{Traces}
\label{sec:traces}

The initial traces for the four terms, constituting the direct-direct part of
the cross section, are given by
\begin{equation}
    \mcM^{\text{nndd}} = \frac{1}{8m^4} \Tr\bigg\{
    \rho_{-p_2} K_{-p_2p_3}^\varkappa (\phi_2) 
    (\gamma \tilde{p}_3 + m) K_{p_3p_1}^\lambda (\phi_1)
    \rho_{p_1} K_{p_1p_3}^\mu(\phi_1') 
    (\gamma \tilde{p}_3 + m) K_{p_3,-p_2}^\nu (\phi_2') \bigg\} 
    g_{\varkappa \nu} g_{\lambda \mu},
\end{equation}
\begin{equation}
    \mcM^{\text{nidd}} = \frac{1}{2m^2+s(\phi)} \Tr\left\{ 
    \rho_{-p_2} K_{-p_2p_3}^\varkappa(\phi_2) 
    (\gamma \tilde{p}_3 + m) K_{p_3p_1}^\lambda(\phi_1)
    \rho_{p_1} \frac{K_{p_1,-p_2}^{\mu \nu}(\phi_1',\phi_2')}{2p_3^+} \right\} 
    g_{\varkappa \nu} g_{\lambda \mu},
\end{equation}
\begin{equation}
    \mcM^{\text{indd}} = \frac{1}{2m^2+s(\phi)} \Tr\left\{ 
    \rho_{-p_2} \frac{K_{-p_2p_1}^{\varkappa \lambda}(\phi_2,\phi_1)}{2p_3^+}
    \rho_{p_1} K_{p_1p_3}^\mu(\phi_1') 
    (\gamma \tilde{p}_3 + m) K_{p_3,-p_2}^\nu(\phi_2') \right\} 
    g_{\varkappa \nu} g_{\lambda \mu},
\end{equation}
\begin{equation}
    \mcM^{\text{iidd}} = - \Tr\left\{ 
    \rho_{-p_2} \frac{K_{-p_2p_1}^{\varkappa \lambda}(\phi_2,\phi_1)}{2p_3^+}
    \rho_{p_1} \frac{K_{p_1,-p_2}^{\mu \nu}(\phi_1',\phi_2')}{2p_3^+} \right\} 
    g_{\varkappa \nu} g_{\lambda \mu}
\end{equation}
[note that $\phi_1'=\phi_2'$ and $\phi_1=\phi_2$ for the ``nidd'' and ``indd''
terms, respectively, see Eqs. (\ref{eq:mm_nidd}) and (\ref{eq:mm_indd}), and
both relations are valid for the ``iidd'' term, see \eqref{eq:mm_iidd}]. In
principle, the traces can be evaluated with the use of the standard techniques
\cite{berestetskii_qed_1982}. Alternative approaches have also been suggested
\cite{hartin_fierz_2016, bragin_dissertation_2019}. The results can be written
in a manifestly Lorentz-invariant form (note that the ``+'' component of a 
four-momentum $p^\mu$ is given by $p^+=k_0p/m$)~\cite{bragin_dissertation_2019},
\begin{align}
    \label{eq:trace_nndd_final}
    \mcM^{\text{nndd}} = & \frac{-1}{4p_2^+p_1^+p_3^{+2}m^4}
    \left[\frac{1}{2} (p_1^{+2} + p_3^{+2}) \Delta_1^2 
    + 2k_1^+p_1^+k_1Z_1 - 2p_1^+p_3^+m^2\right]
    \left[ \frac{1}{2} (p_2^{+2} + p_3^{+2}) \Delta_2^2 
    - 2k_2^+p_2^+k_2Z_2 + 2p_2^+p_3^+m^2\right]
    \nonumber\\&
    - \frac{k_2^+k_1^+}{4p_3^{+2}m^2}
    \Delta_1\Delta_2
    - \frac{(p_1^+ + p_3^+)(p_3^+ - p_2^+)}{4p_2^+p_1^+p_3^{+2}m^4}
    \bigg[ \Delta_1 \Delta_2 
    \left(k_2^+k_1^+Z_1Z_2 - k_2^+p_1^+k_1Z_2
    + k_1^+p_2^+k_2Z_1 - p_2^+p_1^+k_2k_1 \right)
    \nonumber\\&\mspace{110mu}
    + k_2^+p_1^+ k_1 \Delta_2 \Delta_1 Z_2 
    - k_1^+p_2^+ k_2 \Delta_1 Z_1 \Delta_2
    - k_2^+k_1^+ \Delta_1 Z_2 Z_1 \Delta_2 
    + p_2^+p_1^+ k_1 \Delta_2 k_2 \Delta_1\bigg],
\end{align}

\begin{align}
    \label{eq:trace_nidd_final}
    \mcM^{\text{nidd}}
    = \frac{2}{p_3^{+2}[2m^2+s(\phi)]}
    \Bigg[ &m^2(p_3^+ - k_1^+)(p_3^+ + k_2^+)
    + \frac14 (p_1^+ + p_3^+)(p_3^+ - p_2^+) \Delta_1 \Delta_2
    \nonumber\\&
    - \frac12 k_2^+ (p_1^+ + p_3^+) \Delta_1 Z_2
    - \frac12 k_1^+ (p_3^+ - p_2^+) Z_1 \Delta_2
    \nonumber\\&
    - \frac12 p_2^+ (p_1^+ + p_3^+) k_2 \Delta_1
    + \frac12 p_1^+ (p_3^+ - p_2^+) k_1 \Delta_2
    \nonumber\\
    &+ k_2^+k_1^+ Z_1 Z_2 - k_2^+p_1^+ k_1 Z_2
    + k_1^+p_2^+ k_2 Z_1 - p_2^+p_1^+ k_2 k_1 \Bigg],
\end{align}

\begin{equation}
    \label{eq:trace_indd_iidd_final}
    \mcM^{\text{indd}} = \mcM^{\text{nidd}}
    \big|_{\Delta_1^\mu \to -\Delta_1^\mu,
    \Delta_2^\mu \to -\Delta_2^\mu},
    \qquad
    \mcM^{\text{iidd}} = \frac{2p_2^+p_1^+}{p_3^{+2}},
\end{equation}
where
\begin{equation}
    \Delta_1^\mu = \Delta_{p_1}^\mu(\phi_1',\phi_1),
    \quad
    Z_1^\mu = Z_{p_1}^\mu(\phi_1',\phi_1),
    \quad
    \Delta_2^\mu = \Delta_{-p_2}^\mu(\phi_2,\phi_2'),
    \quad
    Z_2^\mu = Z_{-p_2}^\mu(\phi_2,\phi_2'),
\end{equation}
with
\begin{equation}
    \Delta_p^\mu(\phi,\phi') = \pi_p^\mu(\phi) - \pi_p^\mu(\phi'),
    \quad
    Z_p^\mu(\phi, \phi') 
    = \left[ \pi_p^\mu(\phi) + \pi_p^\mu(\phi') \right] / 2.
\end{equation}
In \eqref{eq:trace_nndd_final} a combination of four four-vectors stands for the
product of two scalar products, e.g., $k_1 \Delta_2 k_2 \Delta_1 = (k_1 \cdot
\Delta_2) (k_2 \cdot \Delta_1)$ and analogously for the other combinations.

The results in Eqs. (\ref{eq:trace_nndd_final}), (\ref{eq:trace_nidd_final}),
and (\ref{eq:trace_indd_iidd_final}) can be cast into a more convenient form
with the use of momentum relations for the dressed momenta. First, we notice
that, since ``+'' and ``$\perp$'' momentum components are conserved in the plane
wave, the relations
\begin{equation}
    [\pi_{p'}(\phi) + k - \pi_{p}(\phi)]^{(+,\perp)} = 0
\end{equation}
hold, where $\pi_{p}^\mu(\phi)$ is the fermion four-momentum which comes into
the point $x^\mu$, and $k^\mu$ and $\pi_{p'}^\mu(\phi)$ are the photon and
fermion outgoing four-momenta, respectively. For an analogous combination of the
``--'' components, for each vertex we have the relation
\begin{equation}
    \int \dd x^+ [\pi_{p'}^-(\phi) + k^- - \pi_{p}^-(\phi)] \ee^{i \Phi(x^+)}
    = -i \int \dd x^+ \partial_{+}[\ee^{i \Phi(x^+)}],
\end{equation}
where $\Phi(x^+) = (p'^- + k^- - p^-)x^+ + \mcS_{p'}(\phi) - \mcS_{p}(\phi)$.
Assuming that the boundary terms must not affect observables, we obtain the full
four-momentum conservation law (see Refs.~\cite{mitter_quantum_1975,
mackenroth_nonlinear_2011, ilderton_trident_2011, seipt_two-photon_2012,
hartin_fierz_2016} for similar considerations)
\begin{equation}
    \label{eq:conservation_four-momentum}
    \pi_{p'}^\mu(\phi) + k^\mu - \pi_{p}^\mu(\phi) = 0,
\end{equation}
which, strictly speaking, holds only inside the integral in $x^+$. With the use
of \eqref{eq:conservation_four-momentum}, one can derive the following momentum
relations \cite{bragin_dissertation_2019}:
\begin{equation}
    \label{eq:momentum_relations}
    k \pi_{p'}(\phi) = \frac12 (p^2 - {p'}^2 - k^2),
    \quad
    k \pi_p(\phi) = - \frac12 ({p'}^2 - k^2 - p^2),
    \quad
    \pi_{p'}(\phi) \pi_p(\phi) = -\frac12 (k^2 - {p'}^2 - p^2).
\end{equation}
The relations (\ref{eq:momentum_relations}) allow one to extract instantaneous
parts, i.e., terms $\propto (p_3^2-m^2)$ and $({p_3'}^2-m^2)$ from the ``nndd''
contribution (\ref{eq:trace_nndd_final}) and include them into the ``indd'' and
``nidd'' contributions, respectively. Subsequently, the instantaneous parts can
be extracted from the ``nidd'' and ``indd'' contributions and combined with the
``iidd'' contribution. These rearrangements are significantly simplified if one
employs the coordinate system, defined by
Eqs.~(\ref{eq:canonical_light-cone_basis}) and (\ref{eq:qmu_specified}). The
result is (see Ref.~\cite{bragin_dissertation_2019} for details)
\begin{align}
    \label{eq:trace_nndd_postfinal}
    \tilde{\mcM}^{\text{nndd}} &= -\frac{1}{4p_2^+p_1^+p_3^{+2}m^4} 
    \left[ \frac{1}{2} (p_1^{+2} + p_3^{+2}) \vec{\Delta}_1^{\perp2} 
    + 2p_1^+p_3^+m^2 \right]
    \left[ \frac{1}{2} (p_{2}^{+2} + p_3^{+2}) \vec{\Delta}_2^{\perp2} 
    - 2p_{2}^+p_3^+m^2\right]
    + \frac{k_2^+k_1^+}{4p_3^{+2}m^2}
    \vec{\Delta}_1^{\perp} \vec{\Delta}_2^{\perp}
    \nonumber\\&
    + \frac{(p_1^+ + p_3^+)(p_3^+ - p_2^+)}{8p_2^{+2}p_1^{+2}p_3^{+2}m^4}
    \Bigg\{ 
    \vec{\Delta}_1^{\perp} \vec{\Delta}_2^{\perp}
    \left[ p_2^{+2}p_1^{+2}\vec{\mcZ}^{\perp2} 
    + k_2^{+2}p_1^{+2} 
    \left( m^2 + \frac{1}{4} \vec{\Delta}_2^{\perp2}\right)
    + k_1^{+2}p_2^{+2} 
    \left( m^2 + \frac{1}{4} \vec{\Delta}_1^{\perp2}\right)\right]
    \nonumber\\
    &\mspace{195mu}
    -2p_2^{+2}p_1^{+2}\vec{\mcZ}^{\perp}\vec{\Delta}_1^{\perp}
    \vec{\mcZ}^{\perp}\vec{\Delta}_2^{\perp} \Bigg\},
\end{align}
\begin{align}
    \label{eq:trace_nidd_postfinal}
    \tilde{\mcM}^{\text{nidd}}
    &= \frac{1}{p_2^+p_1^+p_3^{+2}[2m^2+s(\phi)]}
    \Bigg\{ p_3^{+2}\left[ 2p_2^+p_1^+ + (p_2^++p_1^+)^2 \right]m^2
    + p_3^{+2}p_2^+p_1^+ \left( \vec{\Delta}_2^{\perp2} 
    + \vec{\Delta}_1^{\perp2} \right)
    \nonumber\\
    &\mspace{205mu}
    + \left[ p_2^+p_1^+\vec{\mcZ}^\perp 
    + \frac12 p_3^+(p_1^+ + p_3^+) \vec{\Delta}_1^\perp 
    + \frac12 p_3^+(p_3^+ - p_2^+) \vec{\Delta}_2^\perp \right]^2
    \nonumber\\
    &\mspace{205mu}
    - \left[ \frac12 k_2^+(p_1^+ + p_3^+) \vec{\Delta}_1^\perp
    - \frac12 k_1^+(p_3^+ - p_2^+) \vec{\Delta}_2^\perp \right]^2 \Bigg\},
\end{align}
\begin{equation}
    \label{eq:trace_indd_iidd_postfinal}
    \tilde{\mcM}^{\text{indd}} = \tilde{\mcM}^{\text{nidd}}
    \big|_{\vec{\Delta}_1^{\perp} \to -\vec{\Delta}_1^{\perp},
    \vec{\Delta}_2^{\perp} \to -\vec{\Delta}_2^{\perp}},
    \qquad
    \tilde{\mcM}^{\text{iidd}} = -2,
\end{equation}
where
\begin{equation}
    \vec{\mcZ}^\perp = \frac{k_2^+}{p_2^+}\vec{Z}_2^\perp 
    - \frac{k_1^+}{p_1^+}\vec{Z}_1^\perp.
\end{equation}
Note that the final expressions do not depend on the vector $\vec{k}_1^\perp$.
This facilitates the analytical evaluation of the integral in this variable, as
we have mentioned in the main text.

\section{Integrals for the zero-field limit}
\label{sec:zero-field_integrals}

Here, we present the evaluation of the integrals, given in Eqs.
(\ref{eq:sigma_nndd_final}) and (\ref{eq:sigma_indd_nidd_final}), for the case
of a vanishing laser field. The phase $\Phi_v^{\text{dd}}$ is given in
\eqref{eq:phase_dd_zero-field} [see \eqref{eq:ab_defined} for the definitions of
the quantities $a$ and $b$ used below].

The integral in $\rho$ evaluates to a Bessel function of first kind, in
particular \cite{gradshteyn_table_2007},
\begin{equation}
    \int \frac{\dd \rho}{\rho} 
    \exp\left( \frac{i}{\rho} + \frac{i}{4}a^2v^2\tau^2\rho \right)
    = 2i\pi J_0(av\tau).
\end{equation}
For the integrals in $\tau$, formally, one needs to recover the $i \epsilon$
prescription, in order to make them convergent at infinity. On the other hand,
we can rotate the integration contour clockwise by $\pi/2$ and then make the
replacement $\tau \to -i \tau$, after that the $i \epsilon$ prescription is not
necessary (note that $b>a$). One obtains \cite{gradshteyn_table_2007}
\begin{equation}
    \intlim_0^{\infty} \dd \tau \, \tau J_0(av\tau) 
    \ee^{-ibv\tau} 
    = - \frac{b}{\left(b^2 - a^2\right)^{3/2}v^2},
    \quad
    \intlim_0^{\infty} \dd \tau \, J_0(a\tau) 
    \ee^{-ib\tau} 
    = - \frac{i}{\sqrt{b^2 - a^2}}.
\end{equation}
The integral in $v$ is elementary in the case of a vanishing external field. The
evaluation of the integrals in $k_1^+$ is also straightforward. Afterward, one
needs to express the result in terms of $\mu$, which can be written as
\begin{equation}
    \mu = \frac{(p_2^+ + p_1^+)^2}{4p_2^+p_1^+} (1+\vec{t}^{\perp2}).
\end{equation}
We obtain:
\begin{equation}
    \intlim_0^{p_2^++p_1^+} \dd k_1^+ \frac{b}{\left(b^2 - a^2\right)^{3/2}}
    = \frac{p_2^+ + p_1^+}{4 \mu},
    \quad
    \intlim_0^{p_2^++p_1^+} \dd k_1^+ \frac{1}{\sqrt{b^2 - a^2}}
    = \frac{p_2^+ + p_1^+}{4 \sqrt{\mu(\mu-1)}} 
    \ln\left(\frac{\sqrt{\mu}+\sqrt{\mu-1}}{\sqrt{\mu}-\sqrt{\mu-1}}\right).
\end{equation}
Combining everything together, one recovers the final expressions, presented in
Sec. \ref{sec:zero-field}.

\end{document}